\documentclass[iop]{emulateapj}

\newcommand{\ts}{\thinspace}

\shorttitle{CANDELS: Visual Classification}
\shortauthors{Kartaltepe et al.}
\submitted{Accepted for publication in the Astrophysical Journal Supplement Series}

\begin{document}

\title{CANDELS Visual Classifications: Scheme, Data Release, and First Results}

\author{Jeyhan S. Kartaltepe\altaffilmark{1,2}, 
Mark Mozena \altaffilmark{3}, 
Dale Kocevski \altaffilmark{4}, 
Daniel H. McIntosh\altaffilmark{5}, 
Jennifer Lotz\altaffilmark{6}, 
Eric F. Bell\altaffilmark{7}, 
Sandy Faber\altaffilmark{3}, 
Harry Ferguson\altaffilmark{6}, 
David Koo\altaffilmark{3}, 
Robert Bassett\altaffilmark{8}, 
Maksym Bernyk\altaffilmark{8},
Kirsten Blancato\altaffilmark{2, 9},
Frederic Bournaud\altaffilmark{10},
Paolo Cassata\altaffilmark{11},
Marco Castellano\altaffilmark{12},
Edmond Cheung\altaffilmark{3}, 
Christopher J. Conselice\altaffilmark{13}, 
Darren Croton\altaffilmark{8}, 
Tomas Dahlen\altaffilmark{6}, 
Duilia F. de Mello \altaffilmark{14}, 
Laura DeGroot\altaffilmark{15}, 
Jennifer Donley \altaffilmark{16}, 
Javiera Guedes\altaffilmark{17},
Norman Grogin\altaffilmark{6}, 
Nimish Hathi\altaffilmark{15}, 
Matt Hilton\altaffilmark{18},
Brett Hollon\altaffilmark{5}, 
Anton Koekemoer\altaffilmark{6}, 
Nick Liu\altaffilmark{6,19}, 
Ray A. Lucas\altaffilmark{6}, 
Marie Martig\altaffilmark{8}, 
Elizabeth McGrath\altaffilmark{20}, 
Conor McPartland\altaffilmark{21}, 
Bahram Mobasher\altaffilmark{15}, 
Alice Morlock\altaffilmark{13}, 
Erin O'Leary\altaffilmark{2, 22},
Mike Peth\altaffilmark{23}, 
Janine Pforr\altaffilmark{2},
Annalisa Pillepich\altaffilmark{24}, 
Zachary Rizer\altaffilmark{5}, 
David Rosario\altaffilmark{25},
Emmaris Soto \altaffilmark{14}, 
Amber Straughn \altaffilmark{26}, 
Olivia Telford \altaffilmark{27}, 
Ben Sunnquist\altaffilmark{5},
Jonathan Trump\altaffilmark{28},
Benjamin Weiner\altaffilmark{29}, and
Stijn Wuyts\altaffilmark{30}
}

\altaffiltext{1}{School of Physics and Astronomy, Rochester Institute of Technology, 84 Lomb Memorial Drive, Rochester, NY 14623, USA, jeyhan@astro.rit.edu}
\altaffiltext{2}{National Optical Astronomy Observatory, 950 N. Cherry Ave., Tucson, AZ, 85719, USA}
\altaffiltext{3}{University of California Observatories/Lick Observatory, University of California, Santa Cruz, CA 95064, USA}
\altaffiltext{4}{Department of Physics and Astronomy, University of Kentucky, Lexington, KY 40506, USA}
\altaffiltext{5}{Department of Physics, University of Missouri-Kansas City, 5110 Rockhill Road, Kansas City, MO 64110, USA}
\altaffiltext{7}{Space Telescope Science Institute, 3700 San Martin Drive, Baltimore, MD 21218}
\altaffiltext{7}{Department of Astronomy, University of Michigan, 500 Church St., Ann Arbor, MI 48109}
\altaffiltext{8}{Centre for Astrophysics \& Supercomputing, Swinburne University of Technology, P.O. Box 218, Hawthorn, VIC 3122, Australia}
\altaffiltext{9}{Astronomy Department, Wellesley College 106 Central Street, Wellesley, MA 02481}
\altaffiltext{10}{Laboratoire AIM-Paris-Saclay, CEA/DSM/Irfu - CNRS - Universit\'e Paris Diderot, CE-Saclay, F-91191 Gif-sur-Yvette, France}
\altaffiltext{11}{Aix Marseille Universit\'e, CNRS, LAM (Laboratoire dÕAstrophysique de Marseille) UMR 7326, 13388, Marseille, France}
\altaffiltext{12}{IINAFÐOsservatorio Astronomico di Roma, Via Frascati 33, I-00040, Monteporzio, Italy}
\altaffiltext{13}{School of Physics \& Astronomy, University of Nottingham, Nottingham NG7 2RD}
\altaffiltext{14}{Catholic University of America}
\altaffiltext{15}{Department of Physics and Astronomy, University of California, Riverside, CA 92521, USA}
\altaffiltext{16}{Los Alamos National Laboratory, Los Alamos NM}
\altaffiltext{17}{Institute for Astronomy, ETH Zurich, Wolgang-Pauli-Strasse 27, 8093 Zurich, Switzerland}
\altaffiltext{18}{Astrophysics \& Cosmology Research Unit, School of Mathematics, Statistics \& Computer Science, University of KwaZulu-Natal, Durban 4041, South Africa. }
\altaffiltext{19}{Pioneer High School, 601 W. Stadium Blvd., Ann Arbor, MI 48103}
\altaffiltext{20}{Department of Physics and Astronomy, Colby College, Waterville, ME 04901, USA}
\altaffiltext{21}{Institute for Astronomy, University of Hawaii, 2680 Woodlawn Drive, Honolulu, HI 96822}
\altaffiltext{22}{Department of Physics and Astronomy, Macalester College, 1600 Grand Avenue, Saint Paul, MN 55105, USA}
\altaffiltext{23}{Department of Physics and Astronomy, Johns Hopkins University, 3400 North Charles Street, Baltimore, MD 21218, USA}
\altaffiltext{24}{Harvard-Smithsonian Center for Astrophysics, 60 Garden Street, Cambridge, MA 02138, USA}
\altaffiltext{25}{Max-Planck Institut f\"ur Astronomie, Konigstuhl 17, D-69117, Heidelberg, Germany}
\altaffiltext{26}{NASA Goddard Space Flight Center}
\altaffiltext{27}{Astronomy Department,  3910 15th Ave NE, University of Washington, Seattle, WA 98195, USA}
\altaffiltext{28}{Penn State}
\altaffiltext{29}{Steward Observatory, University of Arizona, 933 North Cherry Avenue, Tucson, AZ 85721}
\altaffiltext{30}{Max-Planck-Institut f\"ur extraterrestrische Physik, Giessenbachstrasse 1, D--85748 Garching bei M\"unchen, Germany}

\begin{abstract}

We have undertaken an ambitious program to visually classify all galaxies in the five CANDELS fields down to $H<24.5$ involving the dedicated efforts of over 65 individual classifiers. Once completed, we expect to have detailed morphological classifications for over 50,000 galaxies spanning $0<z<4$ over all the fields, with classifications from 3-5 independent classifiers for each galaxy. Here, we present our detailed visual classification scheme, which was designed to cover a wide range of CANDELS science goals. This scheme includes the basic Hubble sequence types, but also includes a detailed look at mergers and interactions, the clumpiness of galaxies, $k$-corrections, and a variety of other structural properties. In this paper, we focus on the first field to be completed -- GOODS-S, which has been classified at various depths. The wide area coverage spanning the full field (wide+deep+ERS) includes 7634 galaxies that have been classified by at least three different people. In the deep area of the field, 2534 galaxies have been classified by at least five different people at three different depths. With this paper, we release to the public all of the visual classifications in GOODS-S along with the  Perl/Tk GUI that we developed to classify galaxies. We present our initial results here, including an analysis of our internal consistency and comparisons among multiple classifiers as well as a comparison to the Sersic index. We find that the level of agreement among classifiers is quite good ($>70\%$ across the full magnitude range) and depends on both the galaxy magnitude and the galaxy type, with disks showing the highest level of agreement ($>50\%$)  and irregulars the lowest ($<10\%$). A comparison of our classifications with the Sersic index and rest-frame colors shows a clear separation between disk and spheroid populations. Finally, we explore morphological k-corrections between the V-band and H-band observations and find that a small fraction (84 galaxies in total) are classified as being very different between these two bands. These galaxies typically have very clumpy and extended morphology or are very faint in the V-band.

\end{abstract}

\keywords{ 
Galaxies: evolution ---
Galaxies: high-redshifts --- 
Cosmology: observations
}

\section{Introduction}

Since galaxies were first discovered in the early 20th century, astronomers have used information about their structure and morphology to understand galaxy properties in a larger context. We know that in the local Universe, massive galaxies follow the Hubble sequence and their placement along this sequence (i.e., spiral, elliptical, etc.) is closely correlated with many other galaxy properties, such as stellar mass, color, star formation rate, and the relative dominance of a central bulge \citep[e.g.,][]{1994ARA&A..32..115R}. Using this basic separation of galaxy morphology, along with the various correlated properties, astronomers have been able to  piece together a basic picture of galaxy evolution, connecting blue star forming disk-like galaxies to red and dead elliptical galaxies. 

While this picture of galaxy formation and evolution is supported by galaxy studies in the nearby Universe, we would like to understand how it holds up against observations of galaxies in the early Universe. When did the Hubble Sequence, as we know it, form and how have galaxies changed over time?  Were major mergers more important in the early Universe and how did they affect the morphology of progenitor galaxies? Are the highly irregular, clumpy galaxies we see at high redshift the result of a higher gas fraction in the early Universe? Many of these open questions in galaxy evolution can be addressed by understanding the structure of galaxies at high redshift and how galaxy morphology relates to other properties, such as stellar mass, star formation rate, and AGN content.  

Visual classifications have long been used to study galaxy morphology and structure in the local Universe. Large surveys such as the Sloan Digital Sky Survey \citep[SDSS:][]{2003AJ....126.2081A} and the CFHTLS-Deep Survey have enabled the visual classification of thousands of galaxies in the nearby Universe (e.g., \citealt{2007MNRAS.382.1415S,2010ApJS..186..427N}, Bridge et al. 2010, Atkinson 2013) and studies utilizing these classifications have led to a greater understanding of the various correlations described above. The citizen science project {\it Galaxy Zoo} has extended morphological classifications to the general public and has produced catalogs of nearly one million galaxies from the SDSS (\cite{2008MNRAS.389.1179L,2011MNRAS.410..166L,2013MNRAS.435.2835W}, Darg et al. 2010, Simmons et al. 2014, Willett et al. 2015). Thanks to the various Hubble Space Telescope (HST) deep field surveys (such as the Hubble Deep Field, GOODS, COSMOS, AEGIS, and GEMS), we have been able to probe galaxy morphology to ever increasing redshift. The large optical cameras on HST (WFPC2 and ACS) have enabled detailed morphological studies based on visual classifications for  increasing numbers of objects \citep[e.g.,][]{1998ApJ...499..112B,2000ApJ...536L..77B,2005ApJ...625..621B,2005ApJ...625...23B,2009ApJ...697.1971J,2010ApJ...709.1067B,2010ApJ...721...98K} at intermediate redshifts. However, beyond $z\sim1$, these optical surveys start to probe the rest-frame UV morphologies of galaxies, which are sensitive to the regions of most active star formation. In order to study the overall structure of galaxies, and provide the best comparison to morphologies in the local Universe, we wish to trace the structures where the older stellar populations live. This can best be done in the rest-frame optical using near-infrared imaging at $z > 1$

Until the installation of Wide Field Camera 3 (WFC3) on HST during Servicing Mission 4 (SM4) in 2009, near-infrared imaging surveys were limited to ground-based telescopes (either with wide field cameras or with adaptive optics (AO) for small numbers of objects) or with NICMOS on HST \citep[e.g.,][]{2008ApJ...682..303M,2009ApJ...705L..71K, 2011MNRAS.413...80C}.  The small field of view of NICMOS and AO observations placed practical constraints on the total area coverage and thus numbers of galaxies that these surveys were able to study. WFC3 has enabled detailed investigations of large samples of high redshift galaxies for the first time. Since the first WFC3 surveys began, a number of studies have focused on the morphological properties of samples of tens to hundreds of galaxies at high redshift (e.g.,\citealt{2010ApJ...714L..79C,2011ApJ...743..146C,2011MNRAS.417.2770C, 2011ApJ...735L..22S,2012ApJ...745...85L}; {Morishita, Ichikawa, \& Kajisawa 2104; Glikman et al. 2015).  

We can greatly expand on these morphological studies with larger samples from the Cosmic Assembly Near-Infrared Dark Energy Legacy Survey \citep[CANDELS][]{2011ApJS..197...35G, 2011ApJS..197...36K}. CANDELS provides deep, high resolution near-infrared imaging from WFC3 across five of the most commonly studied deep fields. A key goal of the CANDELS survey is to study the structure and morphology of galaxies at $z=1-3$, a key period of galaxy assembly. A number of studies on the morphological properties of galaxies in the CANDELS fields have already been published (e.,g., \citealt{2012ApJ...753..167B, 2012MNRAS.427.1666B, 2012ApJ...757...23K, 2012ApJ...744..148K, 2012ApJ...753..114W,2013MNRAS.432.2012T,2013MNRAS.433.1185M, 2013arXiv1306.4980L}; Villforth et al. 2014; Trump et al. 2014; Guo et al. 2015; Rosario et al. 2015). Our team has undertaken an ambitious effort to visually classify all CANDELS galaxies brighter than $H=24.5$. Once complete, this will result in detailed morphological classifications for over 50,000 galaxies across all five fields, spanning a wide range in redshift ($0<z<4$) -- the largest such sample of classifications at these redshifts. Two publications \citep{2012ApJ...757...23K, 2012ApJ...744..148K} have already made use of these classifications, investigating the role of mergers among Ultraluminous Infrared Galaxies and X-ray selected AGN, respectively, at $z\sim2$. In this paper, we present our classification scheme and initial results based on the first field that we classified (GOODS-S). With this publication, we are also releasing the full set of classifications for GOODS-S (covering over 7000 galaxies) to the public. At the time of writing, we have completed the classifications for two other fields (UDS and COSMOS), which will be released with future publications.

This paper is organized as follows: Section 2 introduces the CANDELS survey and the data sets discussed in this paper.  In Section 3 we present our visual classification scheme. In Section 4 we discuss our results and various comparisons to test for consistency and in Section 5 we summarize our findings. Throughout this paper we assume a  $\Lambda$CDM cosmology with $\rm H_0=70\ts \rm km\ts s^{-1} \ts Mpc^{-1}$, $\Omega_{\Lambda}=0.7$, and $\Omega_{m}=0.3$. All magnitudes are in the AB system unless otherwise stated.

\section{Observations and Datasets}

\subsection{CANDELS}

CANDELS (PIs Faber \& Ferguson; see Grogin et al. 2011; Koekemoer et al. 2011) is an {\it HST} Multi-Cycle Treasury Program to image portions of five of the most commonly studied legacy fields (GOODS-N: \citealt{2004ApJ...600L..93G}, GOODS-S, COSMOS: \citealt{2007ApJS..172...38S}, UDS: \citealt{2007MNRAS.379.1599L}, and EGS: \citealt{2007ApJ...660L...1D}) with the Wide Field Camera 3 (WFC3) in the NIR. The survey has observed all five fields to 2-orbit depth in F125W ({\it J}-band, 2/3 orbit) and F160W ({\it H}-band, 4/3 orbit) and the central regions of GOODS-N and GOODS-S to 10 orbit depth in these bands as well as F105W ({\it Y}-band). ACS parallel imaging has also been obtained for all of these fields in F814W and F606W. For details on the full CANDELS survey, see \cite{2011ApJS..197...35G}. In addition to the CANDELS observations, a portion of GOODS-S was also observed as a part of the WFC3 Early Release Science \citep[ERS; ][]{2011ApJS..193...27W} campaign in {\it Y}, {\it J}, and {\it H}. While we are classifying galaxies in all of the five CANDELS fields, for this paper, we focus on GOODS-S and include the ERS coverage along with CANDELS for full coverage across the entire field.

The CANDELS observations began in Oct. 2010 and were completed in Aug. 2013. For this paper, we use mosaics at three different depths for comparison. First, we use a uniform 2-orbit depth ({\it J} + {\it H}) mosaic across the full field. This mosaic represents the wide coverage that has been obtained for all five CANDELS fields. For the deep region of GOODS-S, we also use a 4-orbit (available at the beginning of the visual classifications) and the final 10-orbit depth mosaic in order to test the dependence of our classifications on image depth. The original images were reduced and drizzled to a $0.06\arcsec$ pixel scale to create each of the mosaics. The details of the data reduction pipeline are described in \cite{2011ApJS..197...36K}. The WFC3 photometry in both {\it J} and {\it H} band were measured using SExtractor version 2.5.0 \citep{1996A&AS..117..393B} in a `cold mode' setup found to work best for extracting $z\sim2$ galaxies by detecting faint sources as well as optimally deblending and resolving multiple source issues (see, for example, \citealt{2008ApJS..174..136C}). 

\subsection{Ancillary Data and Data Products}

In addition to the CANDELS NIR images and SExtractor catalogs, we also use the optical HST-ACS F606W and F850LP mosaics.  For GOODS-S, these data were already publicly available \citep{2004ApJ...600L..93G}. We use the CANDELS consensus photometric redshift catalog (Dahlen et al. 2013). These photometric redshifts were computed based on the photometry in 14 bands: $U$ (VLT/VIMOS), $BViz$ (HST/ACS), F098M, F105W, F125W, F160W (HST/WFC3/IR), Ks (VLT/ISAAC) and 3.6, 4.5, 5.8, 8.0\ts$\mu$m (Spitzer/IRAC). The PSF-matched photometry catalog \citep{2013ApJS..207...24G} is based on $H$-band detected objects with photometry for all ACS and WFC3 data measured using SExtractor and for the rest of the optical-NIR bands using TFIT \citep{2007PASP..119.1325L}.  These photometric redshifts were used to compute rest-frame magnitudes with the code EAZY \citep{2008ApJ...686.1503B} and the and the templates of \cite{2013ApJS..206....8M}. Stellar masses were also computed at these photometric redshifts using BC03 templates \citep{2003MNRAS.344.1000B}, an exponentially declining star formation history, and  a \cite{2003PASP..115..763C} initial mass function. All of these data can be found in Table 4 of (Santini et al. 2014).

\subsection{Sample Selection and Postage Stamps}

In order to maximize the amount of time spent looking at galaxies that are bright enough to be effectively classified, we settled on a magnitude cut of $H<24.5$  based on classifying a test sample of 100 randomly selected galaxies chosen to sample the full magnitude range of our data. Five people classified all of these objects and we chose $H<24.5$ as our cutoff, because we found that at fainter magnitudes, many of the objects were difficult to classify. For GOODS-S, this magnitude cut results in a final sample of 7634 galaxies across the entire field at 2-epoch depth and 2534 galaxies in the deep region of the field. We have made no cuts based on any other galaxy properties as the values for these properties (such as photometric redshifts and stellar masses) are likely to change in future iterations of the catalogs. We also chose not to make an a priori cut based on redshift so that our final catalogs would cover all redshifts, allowing each user to select the range they are most interested in. As such, our final sample of galaxies spans a wide range in redshift (based on the consensus photometric redshifts described above), from $z=0$ to $z\sim 4$ with $\langle z\rangle = 0.98$ and 3511 galaxies at $z=1-3$ where {\it H}-band CANDELS observations probe the rest frame optical. The redshift and stellar mass distributions of this sample are shown in Figure~\ref{distributions} along with the completeness as a function of mass and redshift. The incompleteness is a result of the magnitude cut used as well as the exclusion of objects around the edges of the mosaic with shallower coverage.

\begin{figure}
\epsscale{1.2}
\plotone{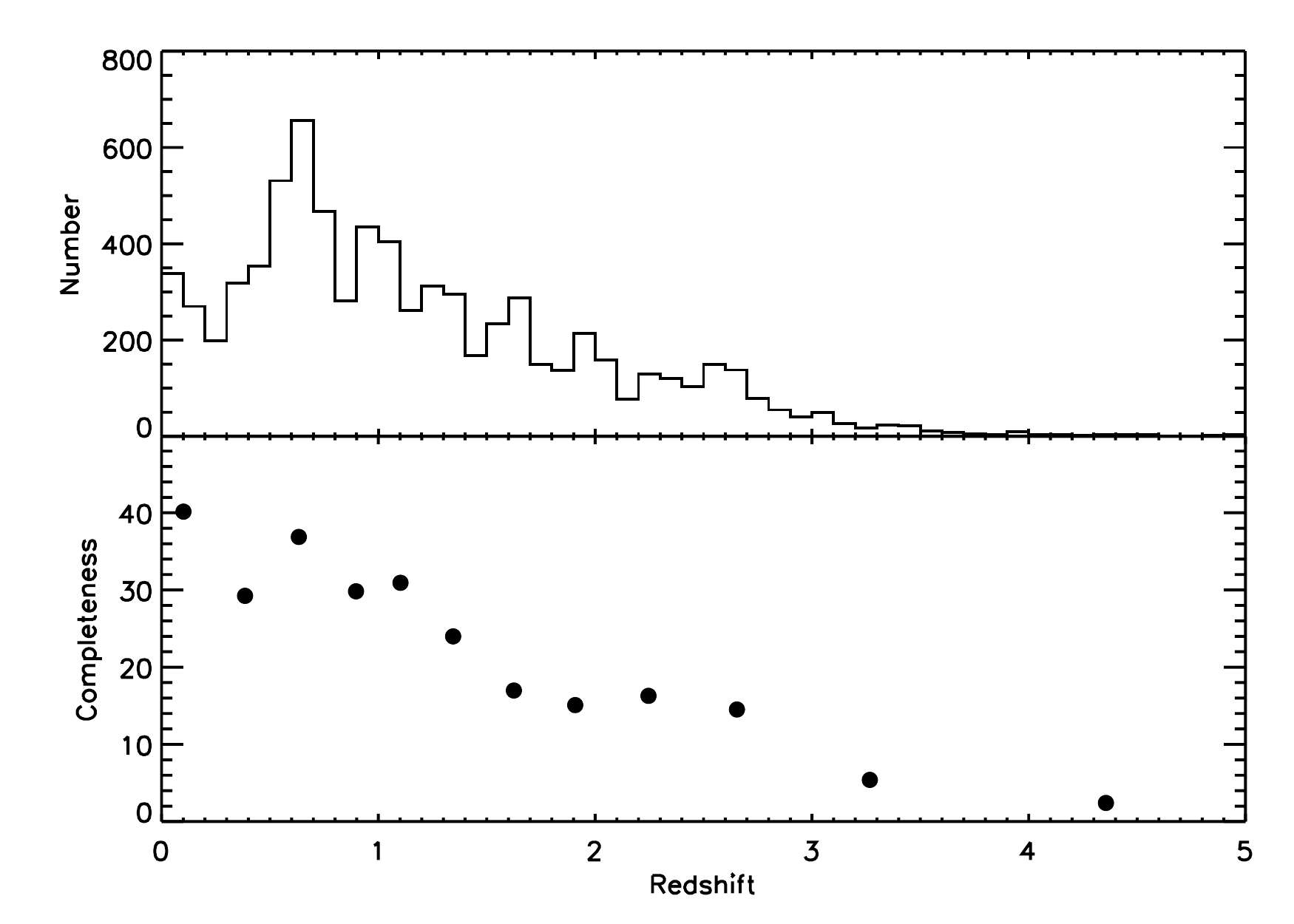}
\plotone{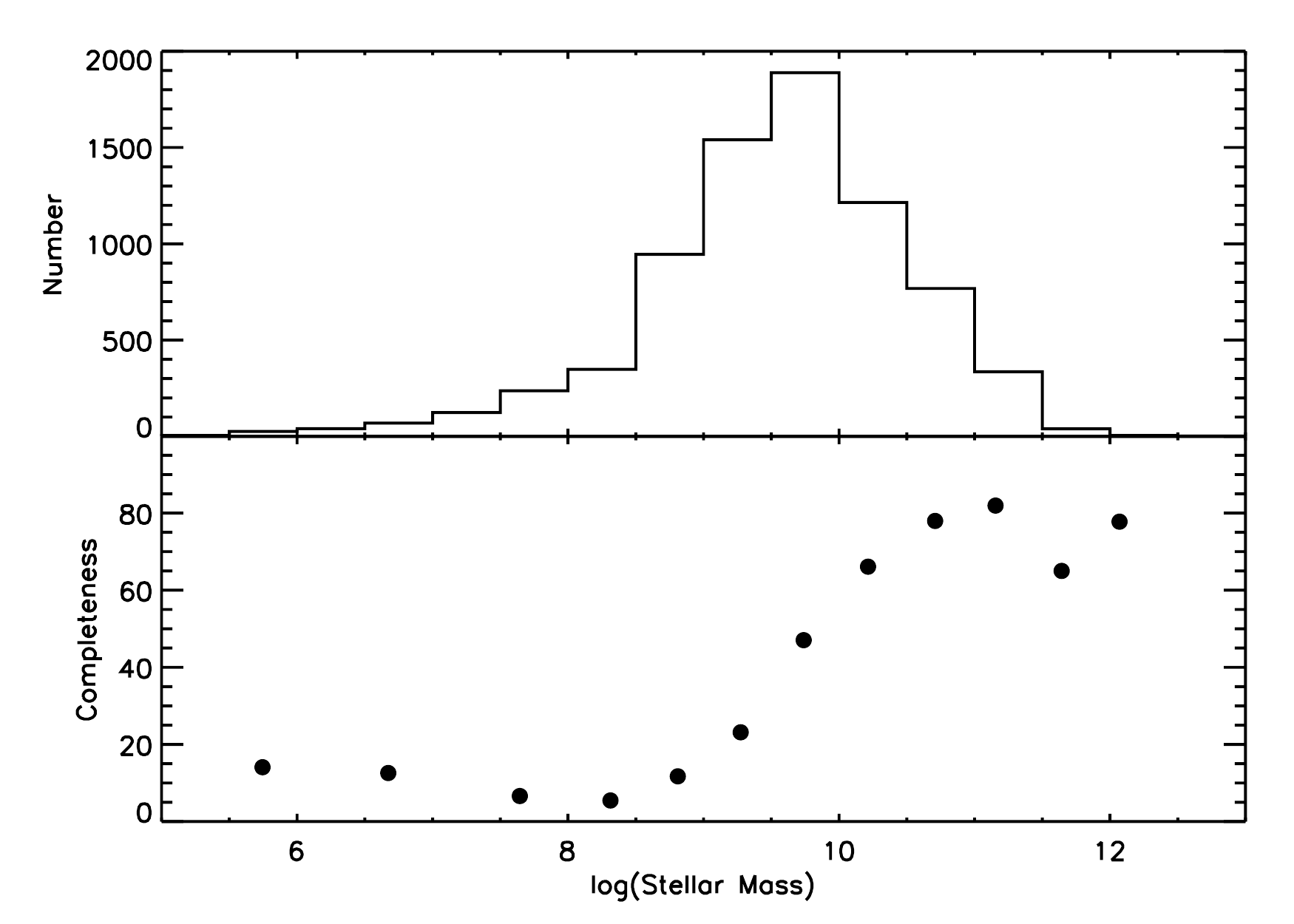}
\caption{Redshift (top) and stellar mass (bottom) distributions and completeness. The completeness indicates the percentage of galaxies in each bin that satisfy the magnitude cut ($H<24.5$) and are observed to at least the 2-epoch depth in the {\it H-}band.}
\label{distributions}
\end{figure}

Following the prescription of \cite{2007ApJS..172..615H}, we used their equations 2 and 3 to scale the size of the postage stamps used for visual classification to the size of the galaxy. We created stamps that were square, with a size equal to the larger of $Xsize$ or $Ysize$ with a minimum size of 84 pixels. In addition, a larger {\it H}-band postage stamp was provided in order to identify potential nearby neighbors. While the visual classifications are conducted primarily in {\it H}-band we also used the optical HST-ACS mosaics to help with classifying and to study the effects of $k$-corrections (see \S 3). We used all four HST bands (F606W, F850LP, F125W, and F160W) as well as the segmentation map from SExtractor, to create cutouts of all 7634 galaxies that we classified. For the deep region of the field, we created three separate sets of cutouts, one using each of the 2-, 4-, and 10- epoch depth mosaics for the F125W and F160W bands. The cutouts for the ACS bands and segmentation map (based off of the initial mosaic of the full field at a non-uniform 6-epoch depth) were the same for each depth.

\section{Visual Classification Scheme}

All of the visual classifications are based primarily on the {\it H}-band WFC3 image, but the {\it J}-band image along with the {\it V} and {\it I}-band ACS images are included to provide additional information and help with the classifications, since the different rest-frame wavelengths and resolutions are sensitive to different structures. In order to determine whether to classify galaxies using all of the bands at once, or to classify each band separately, we conducted a test where two subsets of the co-authors, with five classifiers in each group, visually classified a sample of 100 galaxies -- with one group classifying each band separately and one group classifying based on the {\it H}-band but using the other bands to inform the decision. We then compared the results of these two groups and found that the relative agreement between the groups was rather high but that the group classifying each band separately took much longer and had a difficult time with galaxies that were particularly faint in the optical. Since our ultimate goal is to classify galaxies in all of the CANDELS fields, we opted for the first method to minimize classifier fatigue. In order to study the effects of morphological $k$-corrections, we added a flag (see \S3.3 below) to indicate whether the morphology is different between different bands.

To make the classifications manageable, we divided up all of the galaxies into sets of 200 objects (called a `chunk'). Each classifier was assigned one chunk at a time to classify. Once that chunk was complete, the next one was assigned. Our goal was to have a minimum of three classifiers look at each galaxy so that we could compare the different classifications and look for outliers. For the deep area of the field, we increased that minimum to five classifiers for each galaxy and assigned an independent set of five for each of the three different depths. This means that for the deep region of GOODS-S (2534 galaxies) there are 15 independent sets of classifications, five at each depth.

We developed two different GUIs to allow for a uniform implementation of the classification scheme and to make the classifications go as smoothly as possible. The first is a Perl/Tk based GUI that interacts with the image display tool ds9\footnote{http://hea-www.harvard.edu/RD/ds9/site/Home.html} (shown in Fig.~\ref{gui}). The ds9 window displays all four of the HST bands, in order of increasing wavelength, followed by the segmentation map. This GUI allows for the user to scale each image independently while classifying. The classifier chooses their classification by checking the appropriate boxes and then moves on to the next object. When the set is complete, the GUI writes out a text file with the classifications. We release the software for this GUI (Kocevski 2015) with this paper so that others may use it and our classification scheme for their own galaxy classifications.  The second GUI is a web-based one. Every aspect of this GUI is identical except that the stamps are fixed in scale using arcsinh scaling that was determined to work the best for capturing the range of galaxy features that are present. We allowed the classifiers to choose whichever of the two GUIs they preferred and we note that when we compared the results, we could not find a significant difference based on the GUI chosen. Both of the GUIs have a comment box so that classifiers can note things that do not fall within the classification scheme, problems, or just particularly interesting objects (such as a gravitational lens, for example).

\begin{figure}
\epsscale{1.15}
\plotone{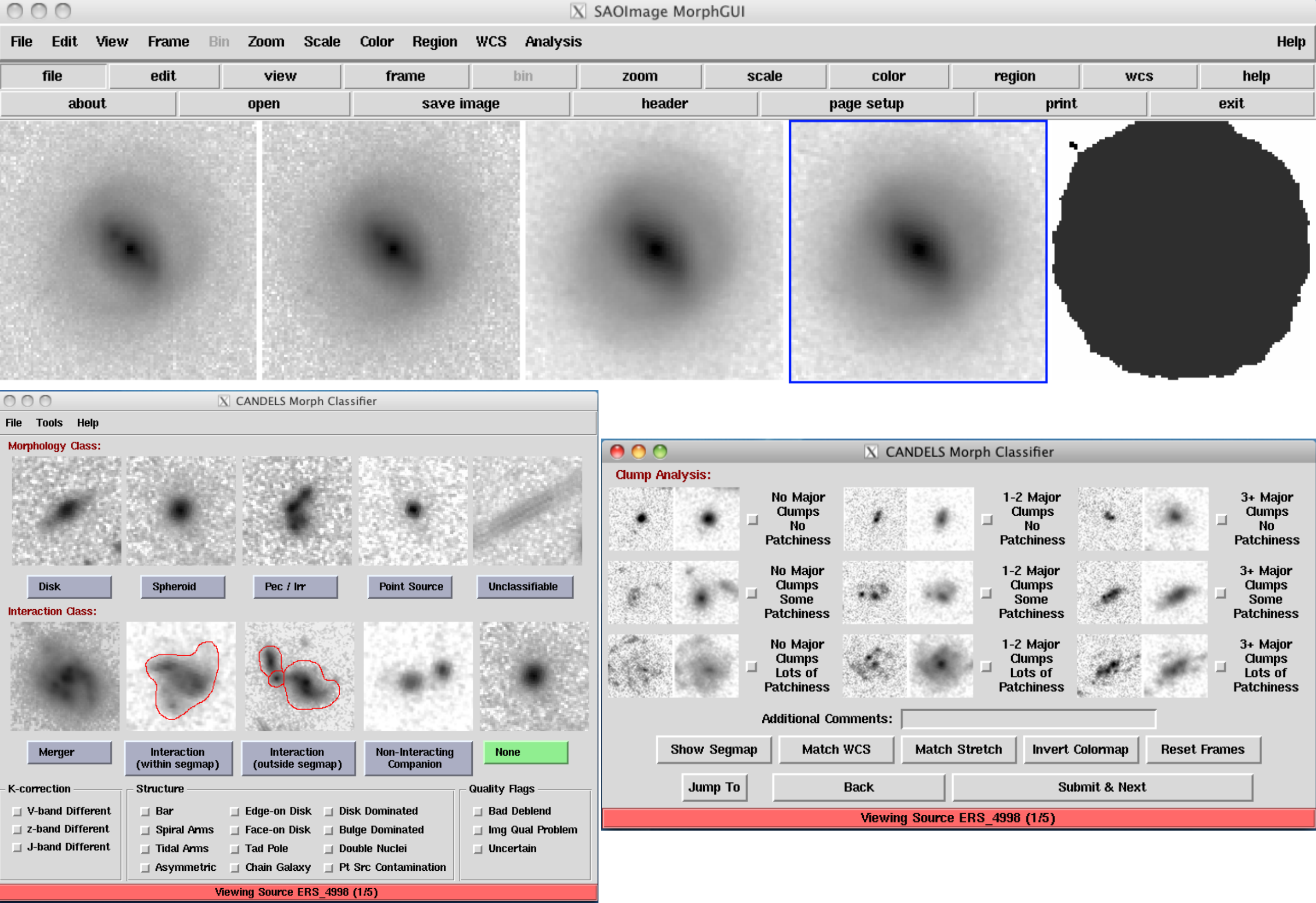}
\caption{Screen shot of the Perl/ds9 GUI used for visual classification of CANDELS galaxies. Top: ds9 window with F606W, F850l, F125W, F160W, and segmentation map images for a sample galaxy. Bottom: GUI window showing visual classification scheme examples and check boxes or the user to select while classifying.}
\label{gui}
\end{figure}

\begin{figure*}
\plotone{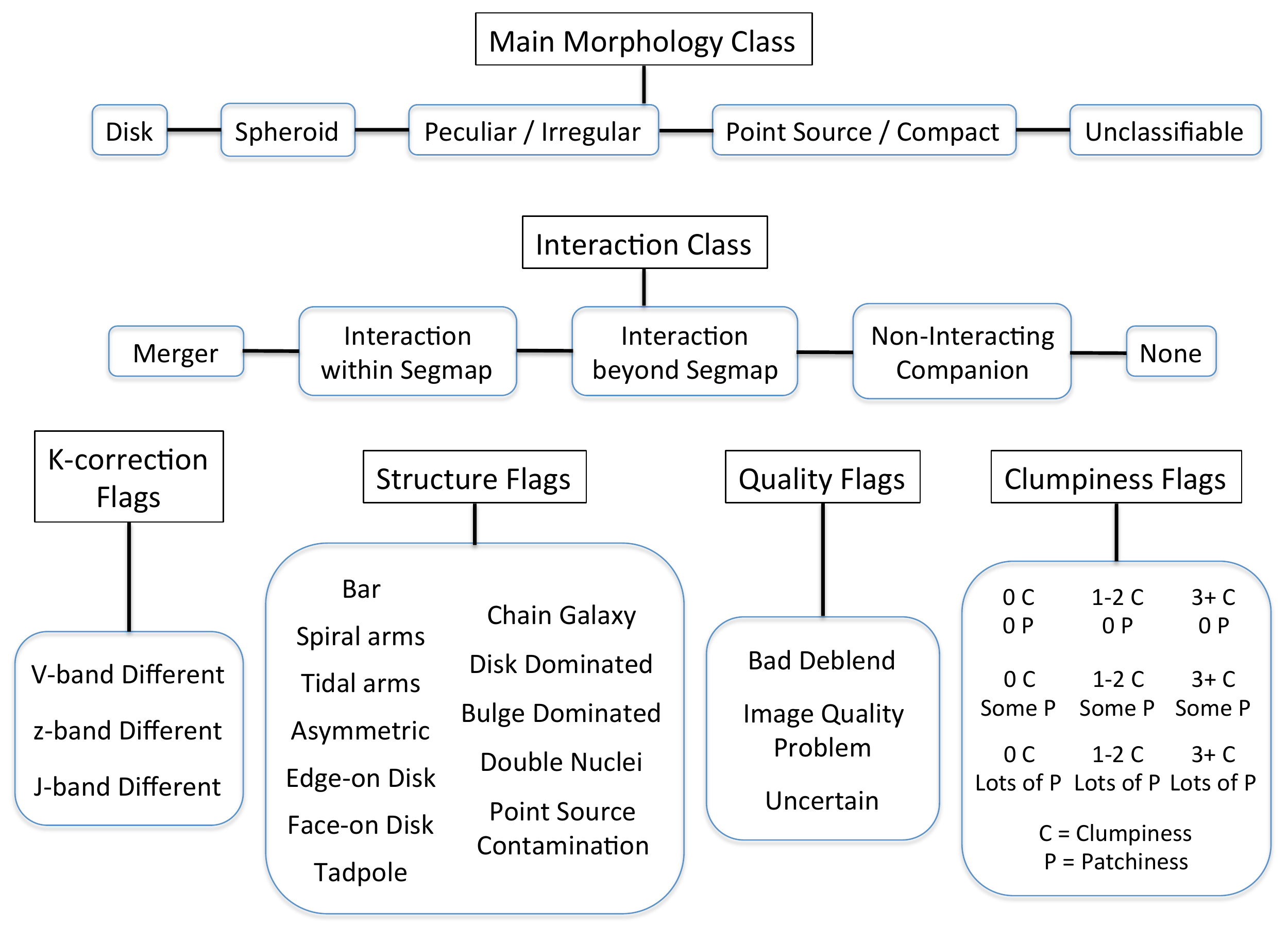}
\caption{Chart illustrating the three different classification levels described in Section 3. At the top level is the main morphology class, where multiple class can be chosen. Next is the Interaction class, only one of which can be chosen. Finally, there are various structural flags, any of which can be chosen as they apply.}
\label{chart}
\end{figure*}

Before beginning the classification process, there are several steps that each classifier must go through. The first is to familiarize themselves with the scheme as described in the following subsections. After this, and after asking any questions they might have, they then go through a training set of 25 galaxies. These 25 galaxies were chosen to reflect the wide range of different classifications that are possible and to illuminate some of the aspects of the scheme that typically create confusion. Throughout the process, we reiterate that there is no right or wrong answer to classification. As long as the scheme is being interpreted correctly, then each classifier's response is valid since they may each see  different aspects of the same galaxy. Once each classifier has gone through these 25 and understand the scheme and why they have made each selection, we assign the first chunk. The first chunk is the same for everybody. This chunk is known as our calibration set. This set of 200 galaxies is simply the first set of objects in the deep area of GOODS-S from the 2-epoch depth mosaic and samples the full magnitude and redshift range of the full sample. By having everyone classify this set first, we can identify any outliers and correct any misunderstandings of the scheme. It also gives the classifiers a chance to get more comfortable with classifying -- the classifiers agree that they approach things a bit differently for the first 20 galaxies in the set than they do for the last 20 in the set. As of the writing of this paper, we have had 65 different classifiers, all of whom have classified this calibration set. This has provided us with a unique sample of 200 galaxies with 65 classifications for comparison -- we will describe these comparisons in detail in \S4.

In the following subsections, we describe each of the three major components of the classification scheme in detail. This scheme is outlined in detail in Figure~\ref{chart}.

\subsection{Main Morphology Class}

The top level of the classification scheme is called the Main Morphology Class. These are based on the typical broad Hubble sequence types and there are five different options to choose from. More than one of these options can be selected for each galaxy, allowing the classifier to indicate intermediate cases. These classes are: 

{\bf 1) Disk:} Disk galaxies have a disk structure that may or may not show clear spiral arms. Disks may also have a central bulge (spheroid component), in which case, the classifier may select both `Disk' and `Spheroid' classes and indicate whether the disk is bulge- or disk-dominated in the structure flags described below.

 {\bf 2) Spheroid:} Spheroid galaxies appear centrally concentrated, smooth, and roughly round/ellipsoidal, regardless of their size, color, or apparent surface brightness. Selecting both `Disk' and `Spheroid' with a bulge-dominated Structure Flag indicates a more early-type galaxy with a modest disk component. 
 
 {\bf 3) Irregular/Peculiar:} These are galaxies that do not easily fall into one of the other categories. This class is meant to indicate galaxies with irregular structure, regardless of surface brightness. This includes objects that are strongly disturbed, such as mergers (see Interaction Classes in the next section) but can also include disks or spheroids that have slightly disturbed morphologies.  For example, an object that has a warped disk or asymmetric spiral arms should have both `Disk' and `Irregular/Peculiar' checked. Or, a spheroidal galaxy with strong asymmetries should have both `Spheroid' and `Irregular/Peculiar' checked.
  
 {\bf 4) Compact/Unresolved:} These objects are either clear point sources, unresolved compact galaxies, or are so small that the internal structure cannot be discerned. A small but clearly resolved spheroidal galaxy should be classified as a spheroid. This class is not meant to be used if the dominant galaxy is another class but has an embedded point source -- there is a Point Source Contamination flag in the Structure flags below.
 
 {\bf 5) Unclassifiable: }These objects are problematic and cannot be classified in any of the other main morphology classes, either because of a problem with the image (e.g., satellite trail, too close to a bright star or galaxy, etc.), the object is not real and should be ignored (e.g., is part of a diffraction spike or is otherwise spurious), or because they are too faint for any structure to be seen.  This class is not meant to be used in combination with any of the other classes. If the object can be classified in any of the bands, that band's classification should be used instead.
 
 \begin{figure*}
\epsscale{1}
\plotone{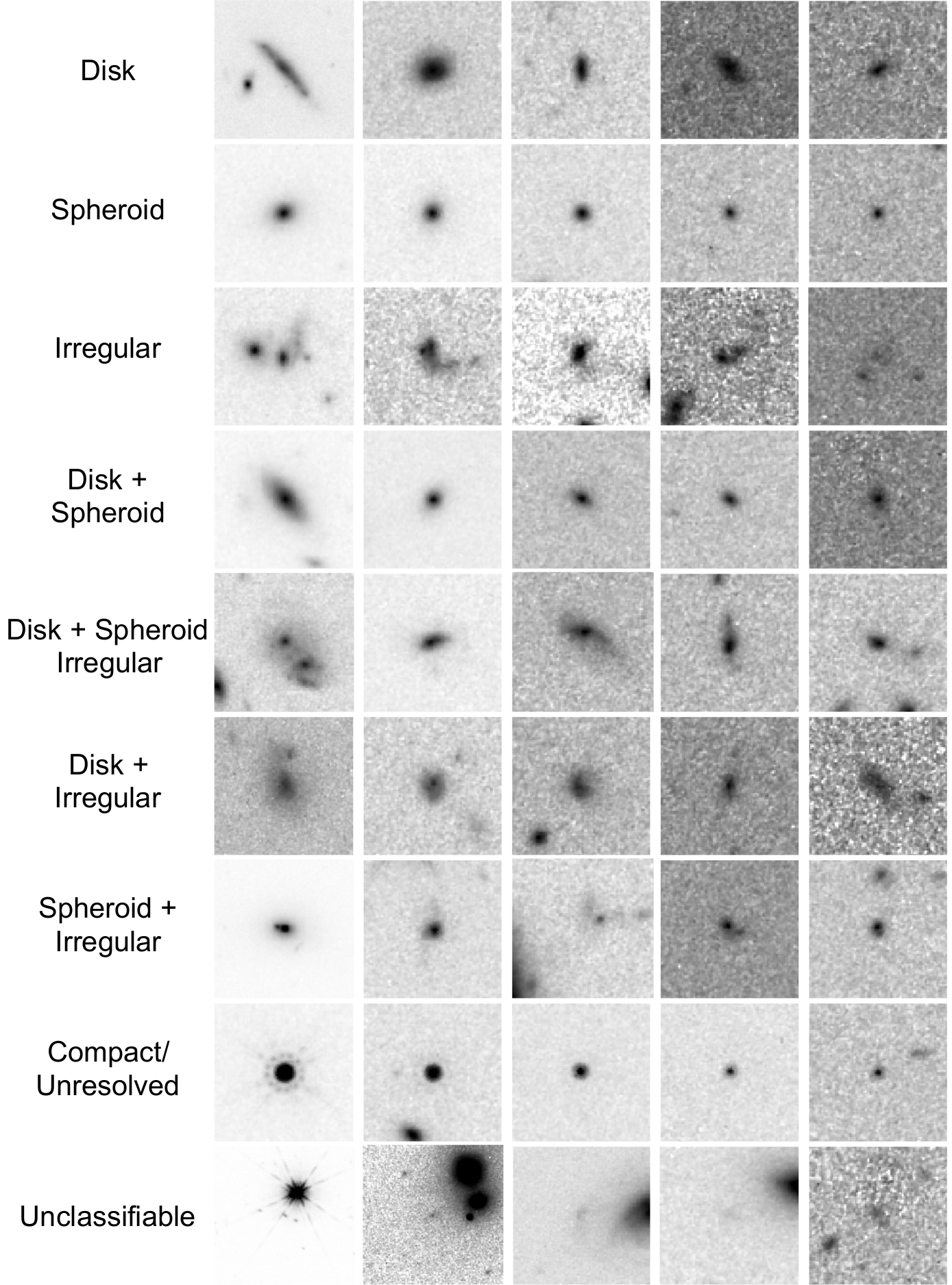}
\caption{Sample HST/WFC3 postage stamps illustrating the main morphology class of the visual classification scheme. The stamps are ordered by {\it H}-band magnitude with the brightest galaxies to the left. The sizes of the stamps follow the prescription described in Section 2.3.}
\label{examples}
\end{figure*}

As noted above, these classes are not mutually exclusive because additional information can be gleaned by choosing more than one class. For example, choosing both disk and spheroid would identify galaxies with both a disk and bulge component. Choosing disk and irregular identifies objects where the disk is still visible but the morphology is slightly disturbed. As a result, there are a total of nine possible combinations of classifications: Disk, Spheroid, Irregular, Disk+Spheroid, Disk+Irregular, Disk+Spheroid+Irregular, Spheroid+Irregular, Compact/Unresolved, and Unclassifiable. Examples of galaxies in these different classes are shown in Figure~\ref{examples}, ordered by their {\it H}-band magnitude. 

\begin{figure}
\epsscale{1.15}
\plotone{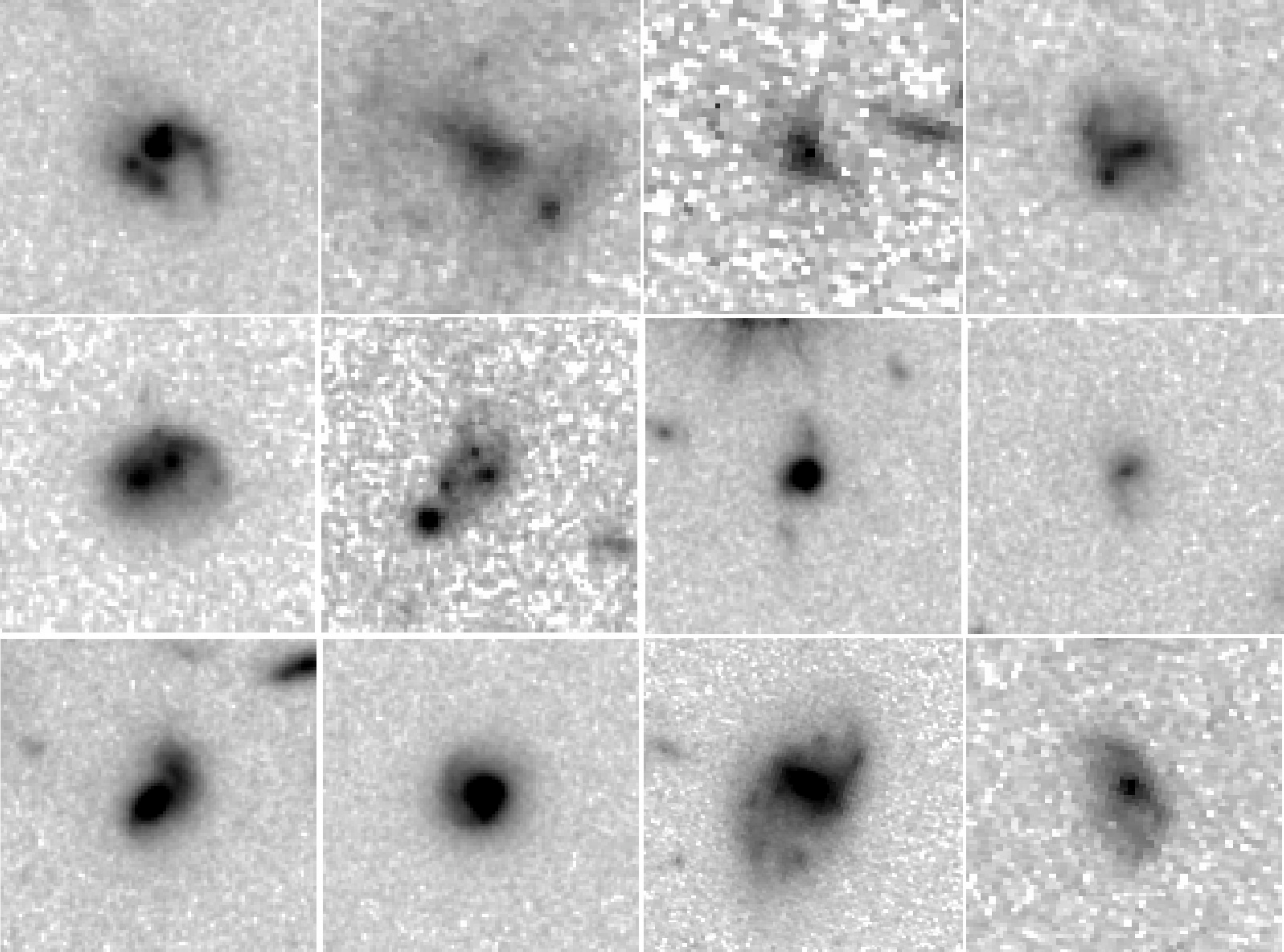}
\caption{Sample HST/WFC3 postage stamps illustrating the different interaction classes. The sizes of the stamps follow the prescription described in Section 2.3.}
\label{mergers}
\end{figure}

\subsection{Interaction Class}

In order to understand the role that galaxy mergers and interactions play in galaxy evolution, we adopted an interaction class that is separate from the main morphology class above. By keeping them separate, the classifier is not forced to choose between a disk and an interaction, for example. These classes are also intentionally kept separate from determining whether or not a galaxy has a clumpy morphology, which is done later on with the clumpiness flags (described in \S3.3). There are four different options for the interaction class and only one of the four (or none) can be selected.

{\bf 1) Merger:} These galaxies are single objects (including sources with double nuclei) that appear to have undergone a merger by evidence of tidal features/structures such as tails, loops, or highly-irregular outer isophotes. All objects classified as mergers should have Irregular/Peculiar selected as one of their main classifications but not all galaxies classified as Irregular/Peculiar are necessarily mergers.

{\bf 2) Interaction within SExtractor segmentation map:}  The primary galaxy appears to be interacting with a companion galaxy within the same {\it H}-band segmentation map. Interactions have clear signatures of tidal interaction (e.g., tidal arms, bridges, dual asymmetries, off-center isophotes, or otherwise disturbed morphologically) -- being apparent close pairs is not enough. To choose interaction over merger, two distinct galaxies must be visible. If more than one companion is present, the classification should be based on the one that appears to dominate the morphology -- usually the larger/brighter one. 

{\bf 3) Interaction beyond SExtractor segmentation map:} The primary galaxy appears to be interacting with a companion galaxy that has its own distinct {\it H}-band segmentation map. By differentiating between interactions within and beyond the segmentation map we can identify galaxies with possible deblending problems. This information is also useful when comparing to various other galaxy properties measured by the CANDELS team, such as photometry, photometric redshifts, stellar masses, and automated classification measures since these will all be based off of the initial source identification done by SExtractor.

{\bf 4) Non-interacting companion:}  These galaxies have a close (visible within the field of view of the large postage stamp) companion (in projection), yet no evidence of tidal interaction or disturbed morphology is apparent. The companion galaxy may be within or beyond the primary galaxy's segmentation map. If each galaxy resides in its own segmentation map, the companion galaxy's segmentation map must be separated from the primary galaxy's segmentation map by less than the diameter of whichever galaxy's segmentation map is larger. Additional information can be used later to determine if the two galaxies are at the same redshift, but either have not yet interacted or lack visible signatures, or if they are simply chance projections. One of the benefits of including this class is that it forces the classifier to consider whether or not they actually see merger/interaction signatures rather than being tempted to call everything with a companion an interaction.

Examples of galaxies in each of the different interaction classes are shown in Figure~\ref{mergers}. 

\subsection{Flags}

We include four different types of flags in our scheme in order to indicate various other structures and features that are not specified in the above morphology and interaction classes.

{\bf Quality flags: } If there are any issues with the images that affect the galaxy or cause the classification to be marked as unclassifiable, then the classifier can choose from three different quality flags. The first of these is `Bad deblend' for cases where the H-band segmentation map has a problem and the galaxy is either over or under deblended. The second quality flag is `Image Quality Problem'. This flag is meant for image problems such as a nearby bright object, the galaxy being too close to the edge of the mosaic, artifacts produced by diffraction spikes or cosmic rays, etc. And finally, there is a flag for `Uncertain' for cases where there are no image quality problems but the classifier is just unsure about their classification.

{\bf K-Correction flags: } These flags are for cases where the difference in morphological structure between the {\it H}-band and any of the bluer bands is so severe that the classifier would select a different classification for that band. The classifier can check any band that meets this condition. This flag should be checked if the object is invisible or substantially fainter in the other bands as well. This flag should not be checked if the differences appear to be solely due to resolution differences. 

{\bf Structure flags:} There is a wide variety of structure flags that can be marked to indicate the presence of interesting/notable features. These are: tidal arms, double nuclei, asymmetric, spiral arms/ring, bar, point source contamination, edge-on disk, face-on disk, tadpole galaxy, chain galaxy, disk-dominated,  and bulge-dominated. 

{\bf Clumpiness/patchiness flags:} Finally, there are a set of flags designed to denote how clumpy/patchy the light distribution of a galaxy is. These flags are set in a $3\times 3$ grid, shown on the right side of Figure~\ref{gui} along with some examples. Clumps are concentrated independent knots of light while patches are more diffuse structures. A central concentration of light is a bulge, not a clump. An object with a continuous surface brightness profile is not considered patchy. Clumps and patches are most clearly seen in the bluer bands so classifiers are asked to look at these for this set of flags.

\begin{figure*}
\epsscale{0.8}
\plotone{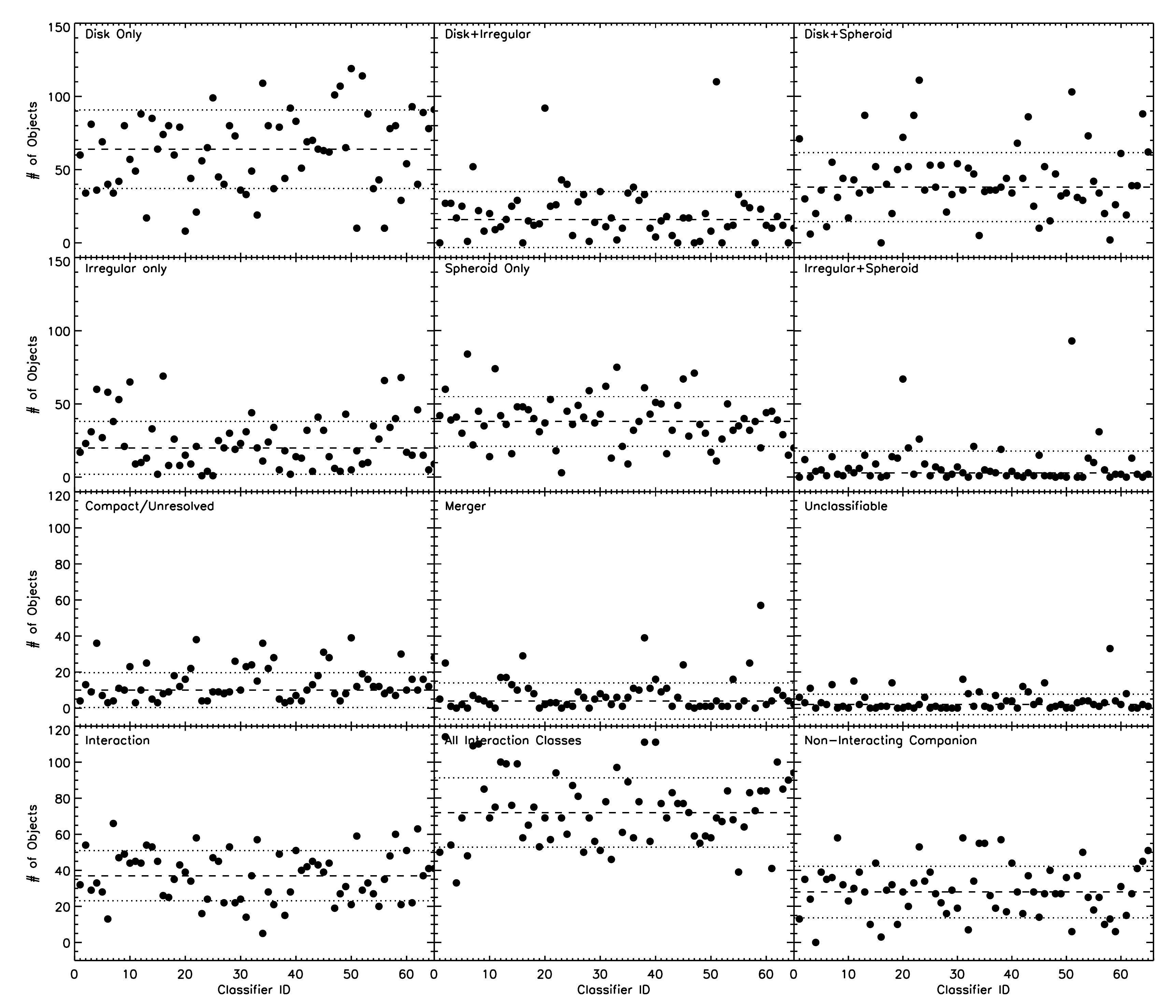}
\caption{Number of objects that each person classified as a given morphological type. The dashed line represents the median number of objects and the standard deviation is shown as dotted lines above and below the median.}
\label{class}
\end{figure*}

\section{Results}

Table~\ref{list} summarizes the various catalogs that are being released with the publication of this paper. There are two different flavors of catalogs -- the `raw' and the `fractional'. The raw catalogs are simply collections of all of the raw classifications. Each object therefore has multiple entries (as many entries as there are classifiers) and each classifier is identified by a unique number. The second set of catalogs, the `fractional' ones, are the ones that are most likely to be useful to the community. These catalogs contain one entry per object and each classification is marked by the fraction of people who checked that box. So if one out of three classifiers classified an object as a disk, one as irregular, and two as a spheroid, then the disk column will have the value 0.33, the spheroid column 0.67, and the irregular column as 0.33. We have created separate catalogs for each depth. There is a 2-epoch depth catalog covering the entire GOODS-S field, one covering just the deep region, and one each covering the deep region at 4- and 10-epoch depths. The calibration set of 200 galaxies classified by all 65 classifiers is included in both 2-epoch depth catalogs. In future papers (M. Mozena, in prep. and D. McIntosh in prep), we will present catalogs that have combined these classifications into other scientifically useful metrics.

\subsection{Calibration Set}

The calibration set of 200 galaxies classified by all 65 of our classifiers provides a unique resource for examining our classification scheme. Figure~\ref{class} shows how often each morphological type is chosen by a given classifier. All nine of the different combinations of main morphology class (except disk+spheroid+irregular since there are very few times all three of these are selected) are shown along with each of the interaction classes. The median number of objects is shown as the dashed line and the standard deviation is shown as the dotted lines above and below the median. This illustrates how likely each classifier is to choose a particular type. For example, we were curious to see if some people were more likely than others to choose `merger'. For the most part, the results are as expected. The disk category is chosen the most often and unclassifiable is rarely selected. One aspect that is immediately noticeable is that there are two people that chose irregular in combination with disk and spheroid often, but those two people are no more likely than others to choose only irregular as a class. Interaction is chosen more often than merger and there are a couple of people more likely to identify an object as a merger than others.

For each of the 200 galaxies in the calibration set, we looked at the overall distribution of classifications. This was an interesting exercise, because while we were looking for evidence of agreement or disagreement, what we found was that the overall distribution of classifications contains useful information. The objects with the highest level of agreement are the simplest cases (for example, the top object in Figure~\ref{dist} is a pure disk with high classifier agreement) while the objects with the lowest level of agreement are the ones with quite complex morphological structure (see object in bottom of Figure~\ref{dist}). Each classifier is seeing a slightly different aspect of the same galaxy when they classify and this extra information is lost if we were to only use one person's classification. 

\begin{figure}
\epsscale{1.2}
\plotone{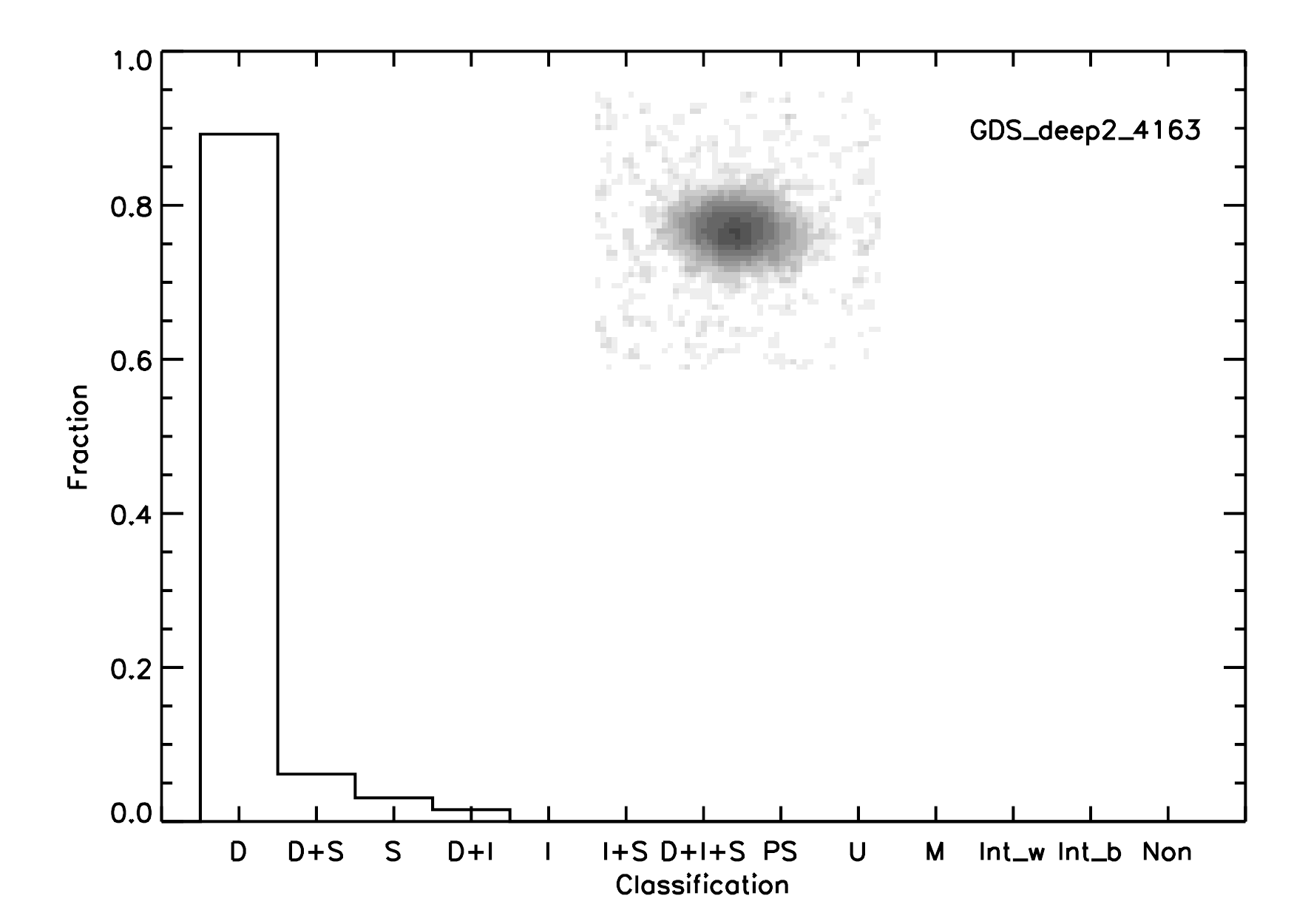}
\plotone{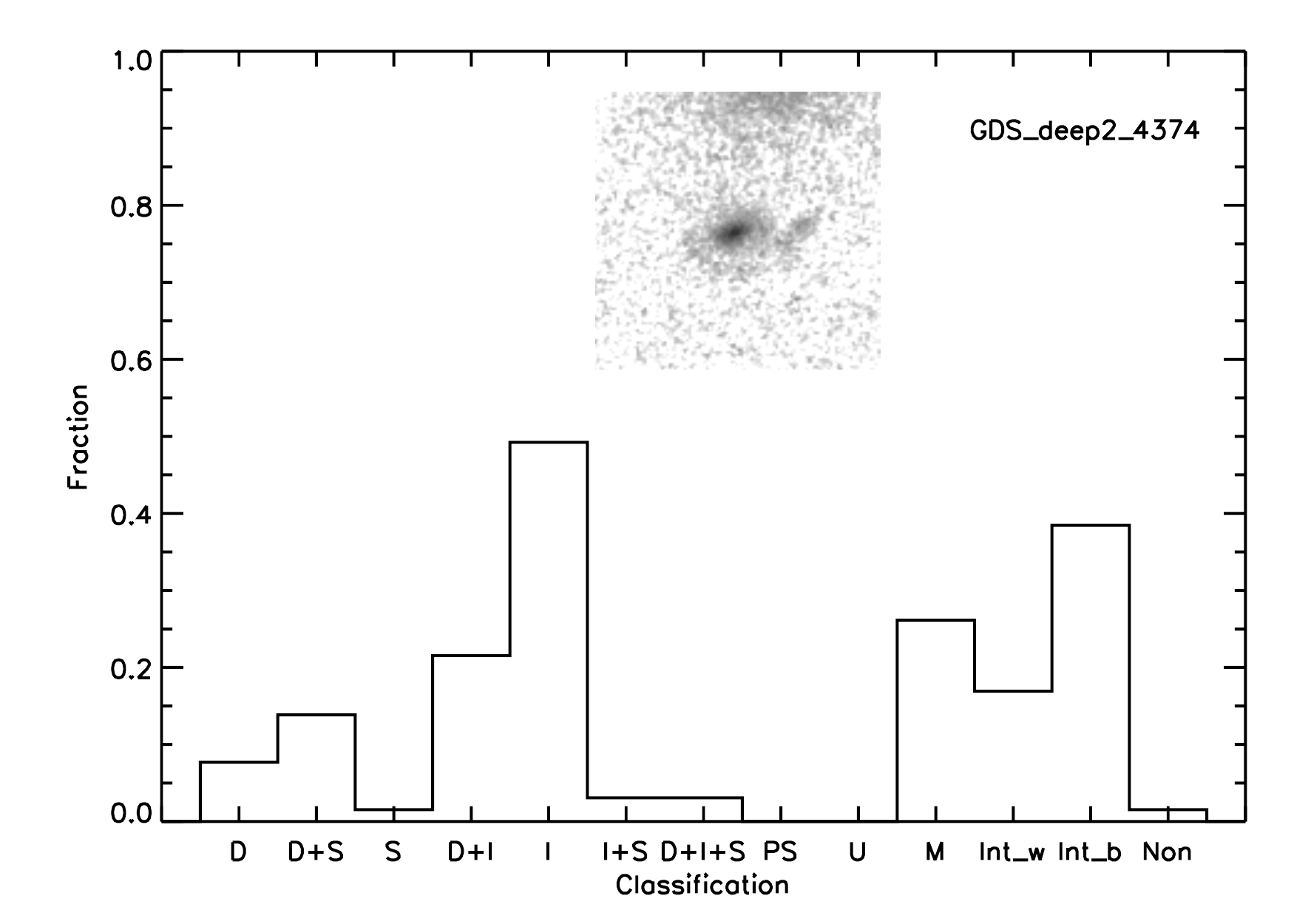}
\caption{The distribution of morphological classifications for two sample galaxies in the calibration set of 200 galaxies classified by 65 people. The top galaxy is one with high agreement where almost everyone agrees it is a disk. A few people also note a bulge component and very few people call it irregular. The bottom galaxy is one with a low level of agreement. The irregular classification has the highest level of agreement, but some people call it a disk and a few see a spheroid component. There is also little agreement about the interaction class (though almost everyone thinks it is merging/interacting on some level).}
\label{dist}
\end{figure}

\subsection{Internal Consistency}

One of the first things to look at with these sets of classifications is internal consistency: How often do classifiers choose the same classifications? Figure~\ref{agree} shows the fraction of objects with high agreement, where high agreement is defined as $>3/5$ of classifiers choosing a single main morphology class, as a function of {\it H}-band magnitude for the 2534 galaxies in the deep area of the field. This fraction is plotted separately for each of the three different depths (with independent classifiers). This plot indicates that the level of agreement is clearly a function of galaxy magnitude, as one would expect. For the brightest galaxies, the agreement is rather high, $\sim 96\%$ of objects have classifications that agree for $>3/5$ of classifiers. The classifications at all three depths agree with each other for these brightest galaxies. The level of agreement stays above $\sim 90\%$ until $H>22$ and then starts to fall off at fainter magnitudes. The dispersion between the different depths also begins to increase. For the faintest galaxies, $H>24$, the level of agreement is lowest for the 2-epoch depth images, but is still above $70\%$ for the 4- and 10-epoch depth images.

\begin{figure}
\epsscale{1.2}
\plotone{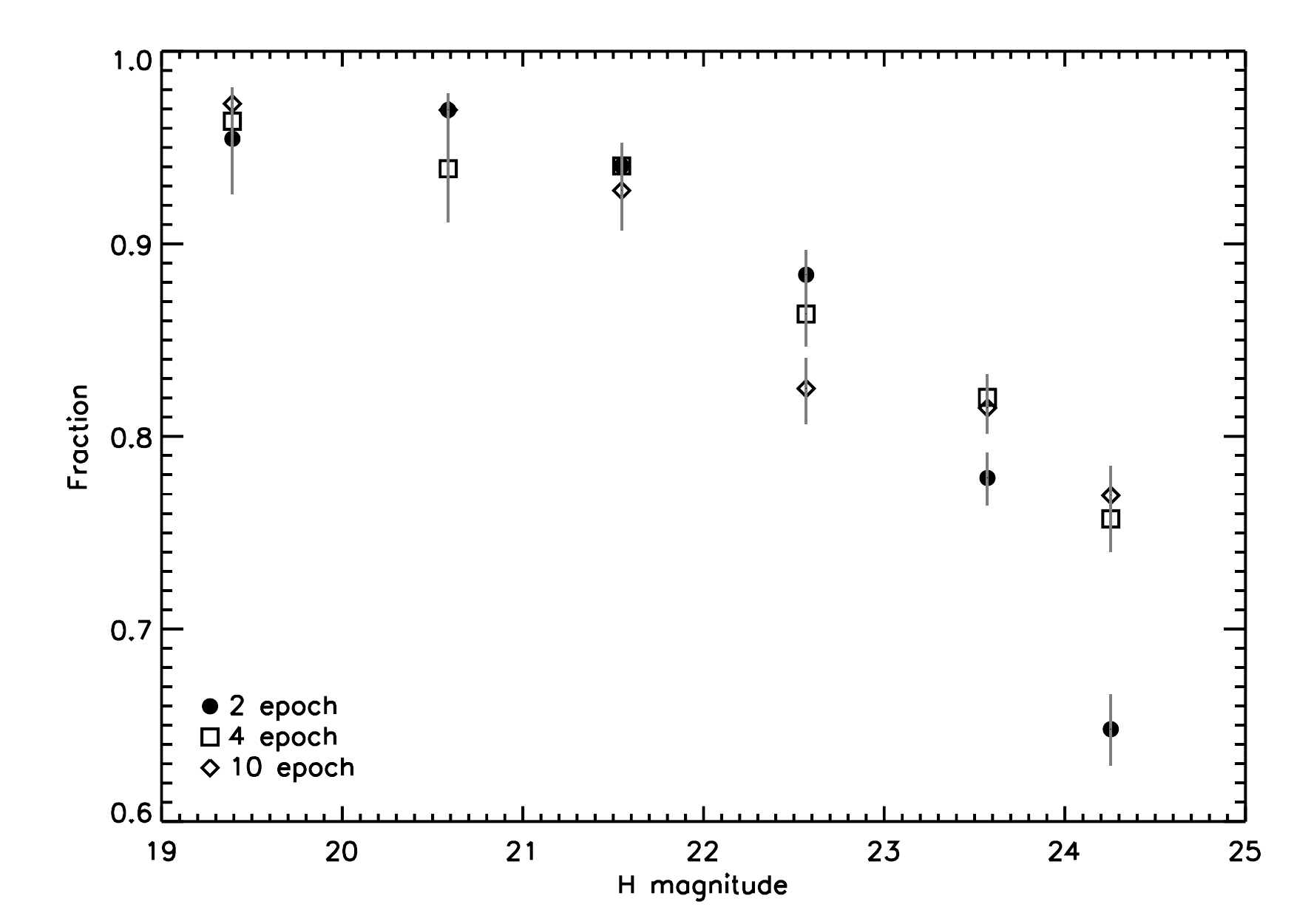}
\caption{Fraction of objects for which $>3/5$ classifiers agree on the classification as a function of {\it H}-magnitude and separated by the depth of the images classified. The error bars on each point reflect the $1\sigma$ binomial confidence limits given the number of objects in each category, following the method of \protect\cite{2011PASA...28..128C}. Even at the faintest magnitudes included in our sample, the fraction of objects with high agreement is still above $\sim70\%.$ }
\label{agree}
\end{figure}

Figure~\ref{agree_sep} shows the same information, but split by the individual main morphology class. Again, a threshold of $>3/5$ classifiers is shown for each type. Since the fractions show how often each class was chosen, the absolute fraction on the y-axis cannot be used to determine the level of agreement, but rather how often a given type was selected with a high confidence level. For example, the fraction of objects classified as disks is high while those classified as irregular are low. The fraction of objects classified as disks with high agreement is consistent across the different depths, until $H>23$, and even then the difference is only slight. However, for spheroids, the difference between the different depths is more pronounced. The classifiers are less likely to agree that an object has a spheroid component in the deepest images, and the fraction of objects with a spheroid component decreases slightly with increasing magnitude. Perhaps this is due to the presence of a disk component becoming easier to see in the deepest images and for brighter galaxies. Overall, there is a low fraction of objects that are classified as irregular with high agreement, but this fraction increases for the faintest galaxies. Classifiers are also more likely to call something irregular for the 2-epoch images than for the deepest images. The consistency of the increasing fraction of irregular galaxies for the 4- and 10- epoch deaths suggests that this is a physical trend and not just the result image depth. The compact/unresolved fraction agrees for all of the depths and the fraction of objects decreases with increasing magnitude. This makes sense since a number of the brightest objects in our sample are point sources. And finally, there is a very low fraction of objects that is classified as `unclassifiable' and this increases only slightly for the faintest galaxies. With this small number of objects, there is no discernible difference between the different depths.

\begin{figure}
\epsscale{1.2}
\plotone{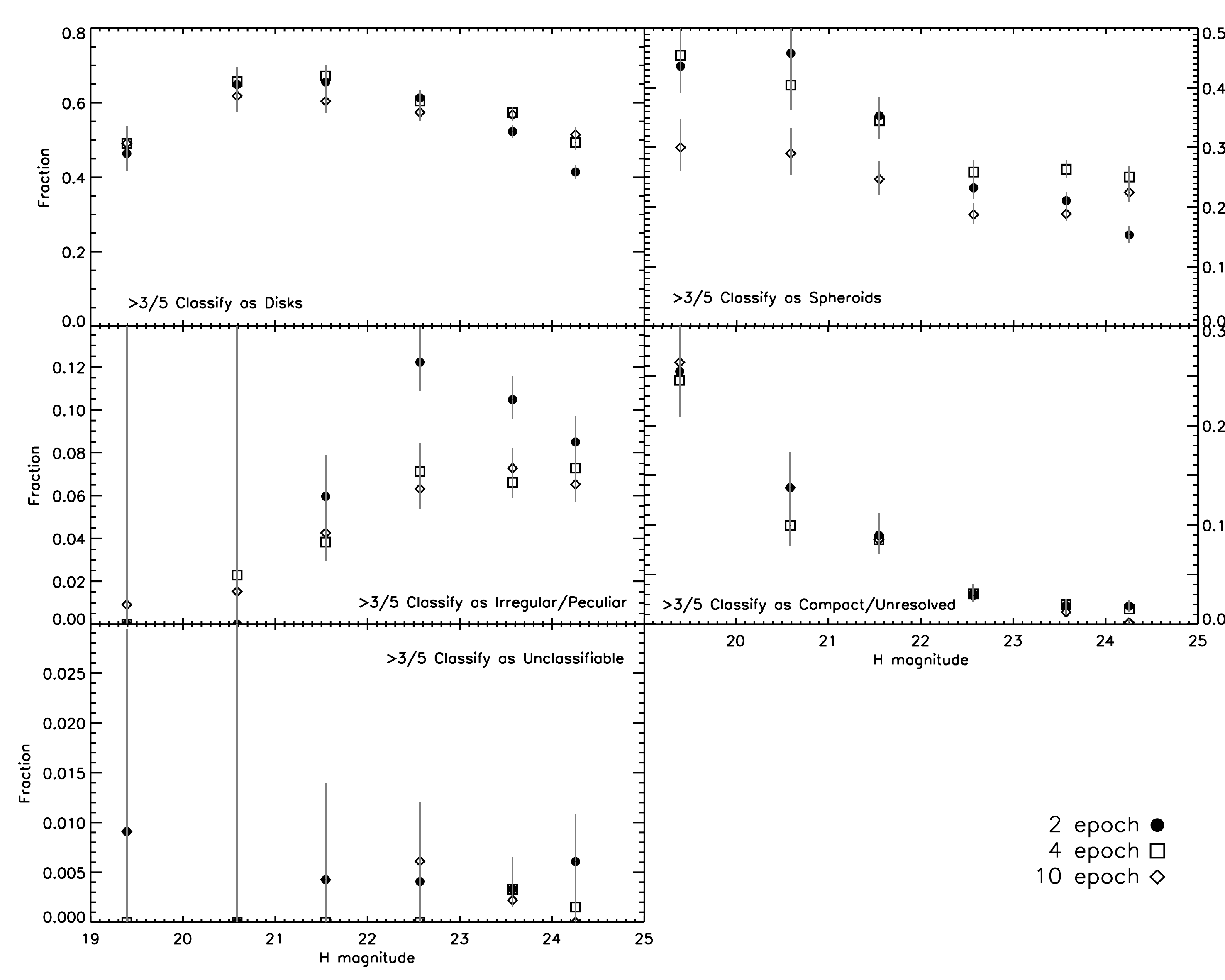}
\caption{Fraction of objects for which $>3/5$ classifiers agree on the classification as a function of {\it H}-magnitude and separated by the depth of the images classified. Each panel represents a separate main morphology class. The error bars on each point reflect the $1\sigma$ binomial confidence limits given the number of objects in each category, following the method of \protect\cite{2011PASA...28..128C}. }
\label{agree_sep}
\end{figure}

Figure~\ref{disagree} shows the fraction of objects for which the classifiers disagree (defined as only one or two of the classifiers choosing a given class) on the particular morphological type of interest. For each main morphology class, the fraction of objects that the classifiers disagree on increases with increasing magnitude. For disks, irregulars, and unclassifiable objects, there is a significant difference between the different depths for the faintest galaxies such that there is a higher fraction of galaxies with disagreement at the 2-epoch depth than for the deeper images.

\begin{figure}
\epsscale{1.2}
\plotone{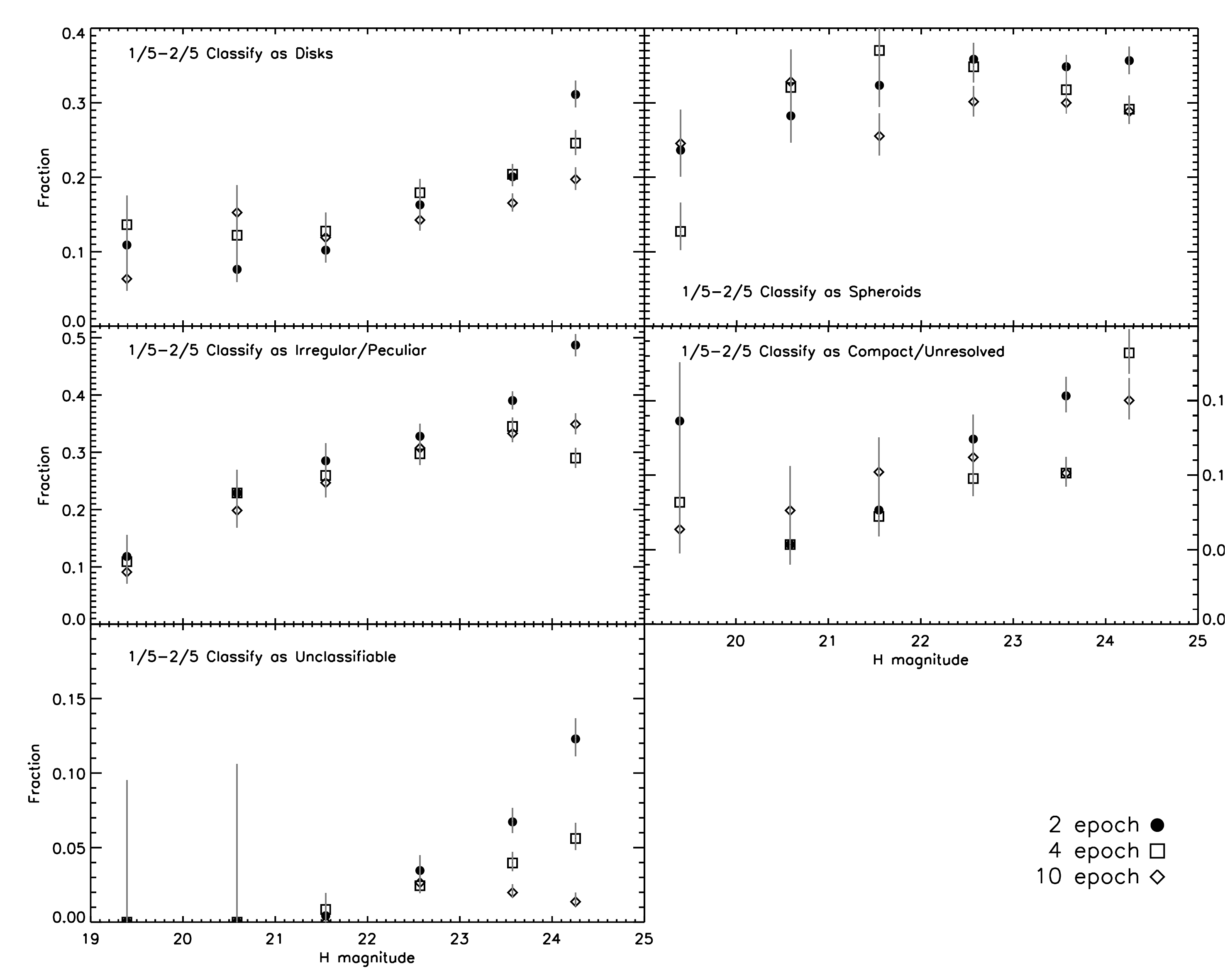}
\caption{Fraction of objects for which $1/5 - 2/5$ classifiers agree on the classification as a function of {\it H}-magnitude and separated by the depth of the images classified. Each panel represents a separate main morphology class. The error bars on each point reflect the $1\sigma$ binomial confidence limits given the number of objects in each category, following the method of \protect\cite{2011PASA...28..128C}.  }
\label{disagree}
\end{figure}

Figure~\ref{delta} illustrates the level of agreement in a different way. Plotted is the difference in the fraction of classifications for each galaxy between the wide (2-epoch) and deep (10-epoch) depths. For example, if for a given galaxy, $3/5$ of classifiers classified it as a disk at both depths, then its value on this plot would be zero. Likewise if $2/5$ or $5/5$ classified it as a disk at both depths. However, if $3/5$ classified it as a disk at the 10 epoch depth and only $1/5$ classified it as a disk at the 2-epoch depth, then its value on this plot would be 0.4. This difference is shown on the top plot for disks, spheroids, and irregulars. All objects with a value of zero have complete agreement between the two different depths for completely independent set of classifiers). This plot illustrates that the level of agreement is highest for disks and the lowest for irregulars. The asymmetry in the distributions for spheroids and irregulars (both have more objects with a negative difference) indicates that a higher fraction of people choose those classifications in the shallower images. The differences for the interaction classes are shown in the bottom plot. Overall, the level of agreement is worse than for the main morphology plot (the distributions are broader). There is the highest agreement for the `any interaction' set, which includes mergers and both interactions within and beyond the segmentation map, as expected. There is a similar level of agreement for the non-interacting companion class. 

\begin{figure}
\epsscale{1.2}
\plotone{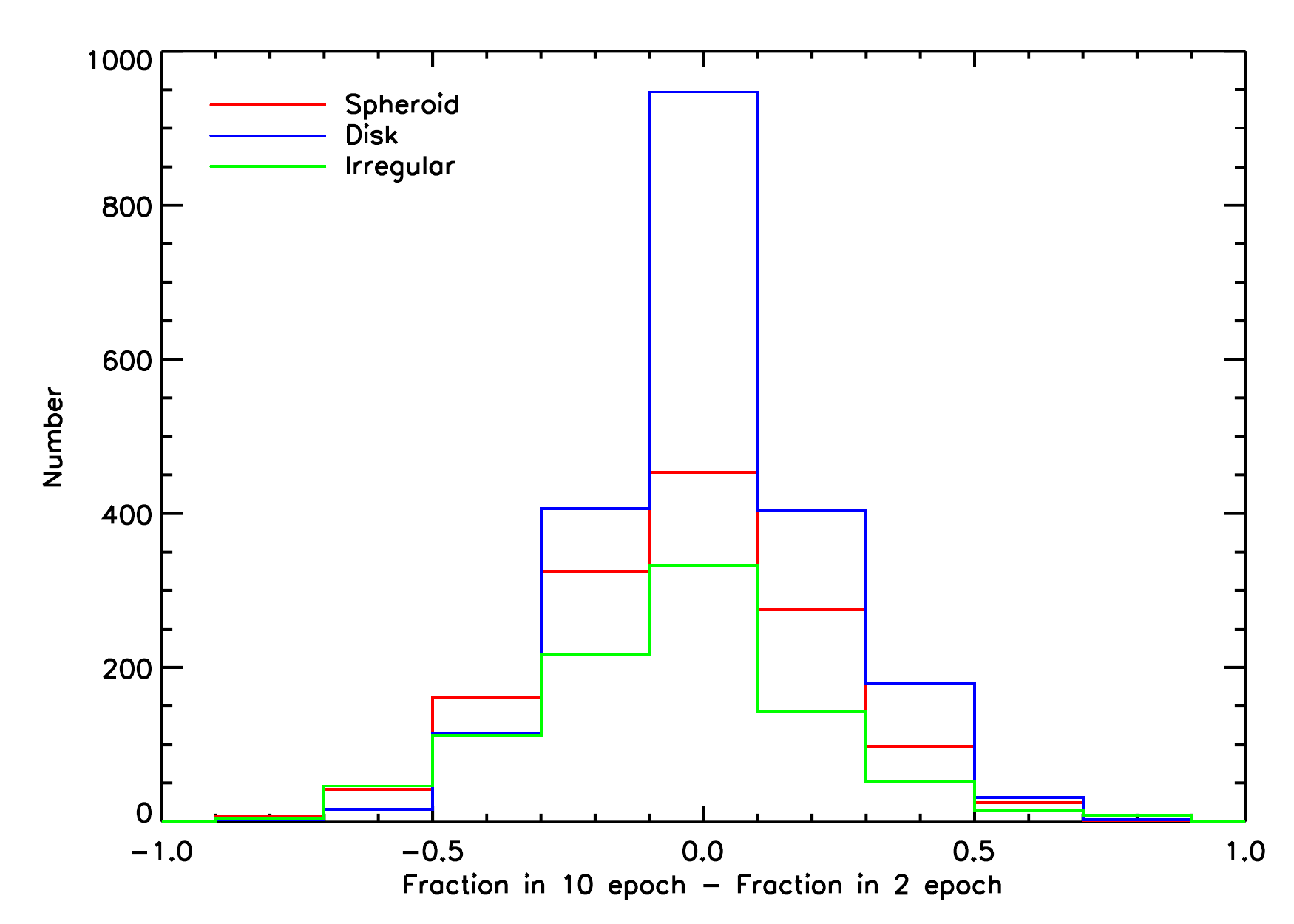}
\plotone{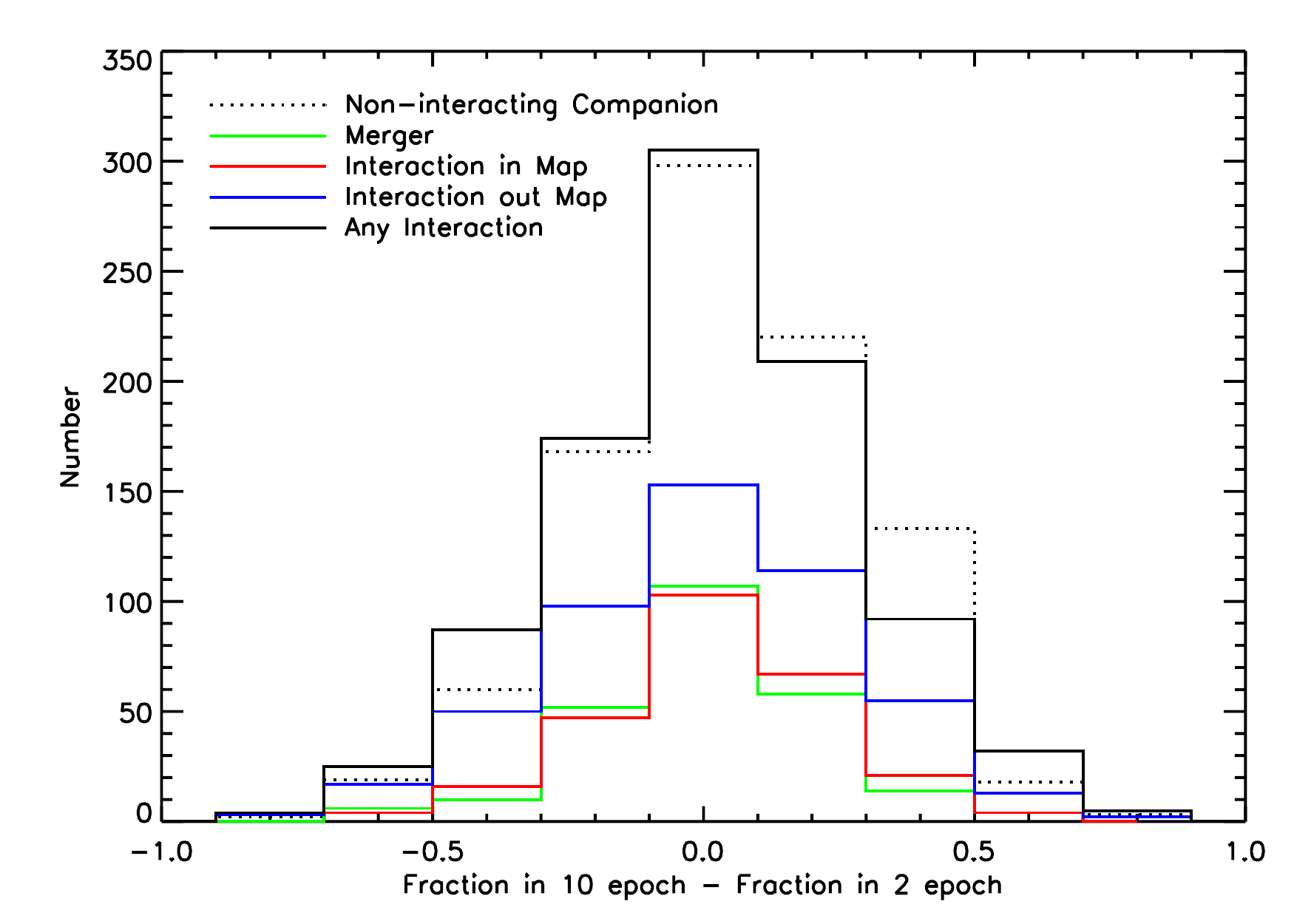}
\caption{The distribution of the differences between fraction of classifiers choosing a given class for a particular object at two different CANDELS depths. Top: Difference for the Disk, Spheroid, and Irregular main morphology classes. Bottom: Difference for each interaction class.}
\label{delta}
\end{figure}

\subsection{Disk/Spheroid Separation}

Ideally, one would like to apply various quantitative measures for galaxy morphology whenever possible. Several methods have been developed for automatically classifying galaxies using image statistics in the local and low redshift Universe, such as CAS \cite[Concentration, Asymmetry, and Clumpiness:][]{1996ApJS..107....1A, 2000ApJ...529..886C, 2003ApJS..147....1C} and Gini/M$_{20}$ \citep{2003ApJ...588..218A, 2004AJ....128..163L}. While some of these methods seem to work well for separating disks from spheroids at low redshifts, they typically do not work as well for identifying mergers and interactions \citep[e.g.,][]{2010ApJ...721...98K} or are only sensitive to mergers at certain stages \citep{2008MNRAS.391.1137L} and they have yet to be thoroughly tested and calibrated at high redshifts with large samples of visually classified galaxies. Our large sample of visually classified galaxies will enable detailed studies of these methods at high redshifts for the first time (e.g., J. Lotz et al., in preparation). The identification of mergers will be the subject of a future paper. Here, we investigate how well our scheme separates disks from spheroids.

First, we compare  our visual classifications to the Sersic index, typically used as a way to separate disks from bulges, using GALFIT \citep{2002AJ....124..266P} measurements for the CANDELS data in GOODS-S \citep{2012ApJS..203...24V}. GALFIT fits the two-dimensional galaxy light profile in an image using a $\chi^{2}$ minimization routine to estimate the best-fit Sersic profile of the galaxy. Reliable fits were obtained for 6225 of our visually classified galaxies in GOODS-S, after excluding objects in noisy areas of the mosaic and objects with unrealistic parameters) . The distribution of Sersic indices ($n$, where $n=0.5$ corresponds to a Gaussian profile, $n=1$ to an exponential profile, and $n=4$ to a de Vaucouleurs profile) are shown in Figure~\ref{galfit}, separated by their main morphology class in the top panel, and their interaction class in the bottom panel.  In the top panel, galaxies are separated by their relative diskiness/bulginess, into `Mostly Disk'  ($>2/3$ classifiers call it a disk and $<2/3$ classifiers call it a spheroid), `Mostly Spheroid' ($>2/3$ classifiers call it a spheroid and $<2/3$ classifiers call it a disk), `Disk + Spheroid' ($>2/3$ classifiers call it a disk and $>2/3$ classifiers call it a spheroid) and `Irregular' ($>2/3$ classifiers call it irregular, $<2/3$ classifiers call it a disk, and $<2/3$ classifiers call it a spheroid). This plot shows a significant difference in the distribution of Sersic indices for each of these classes. The `Mostly Disk' group is narrowly peaked around a value of 1 ($\langle n\rangle=1.01$). The `Mostly Spheroid' group has a broad distribution with $\langle n\rangle=2.98$, and the `Disk + Spheroid' group is in between ($\langle n\rangle=2.53$), as expected. The `Irregular' group, which excludes galaxies that are clearly disks or spheroids more closely resembles the disk distribution with $\langle n\rangle=1.33$.

\begin{figure}
\epsscale{1.2}
\plotone{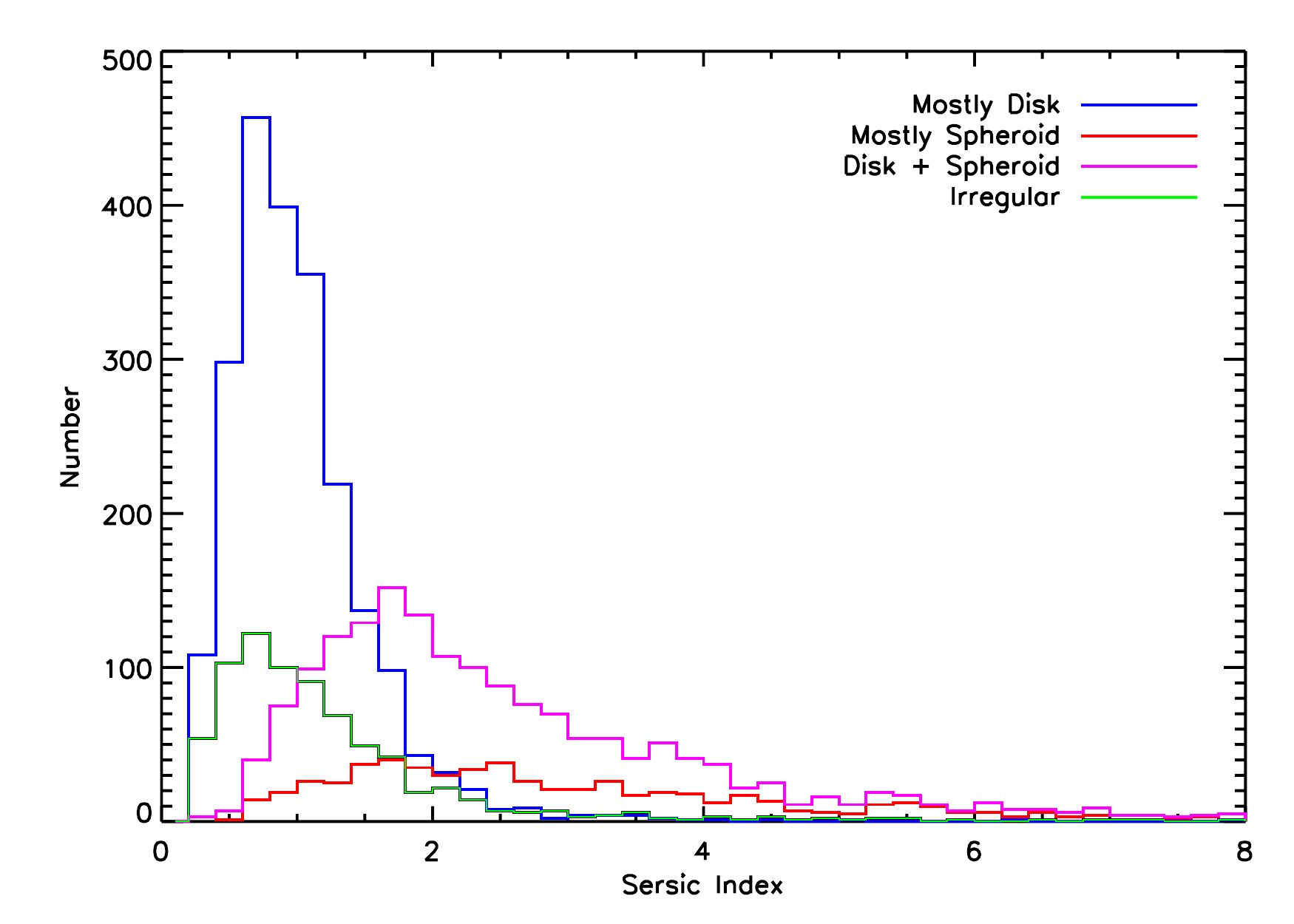}
\plotone{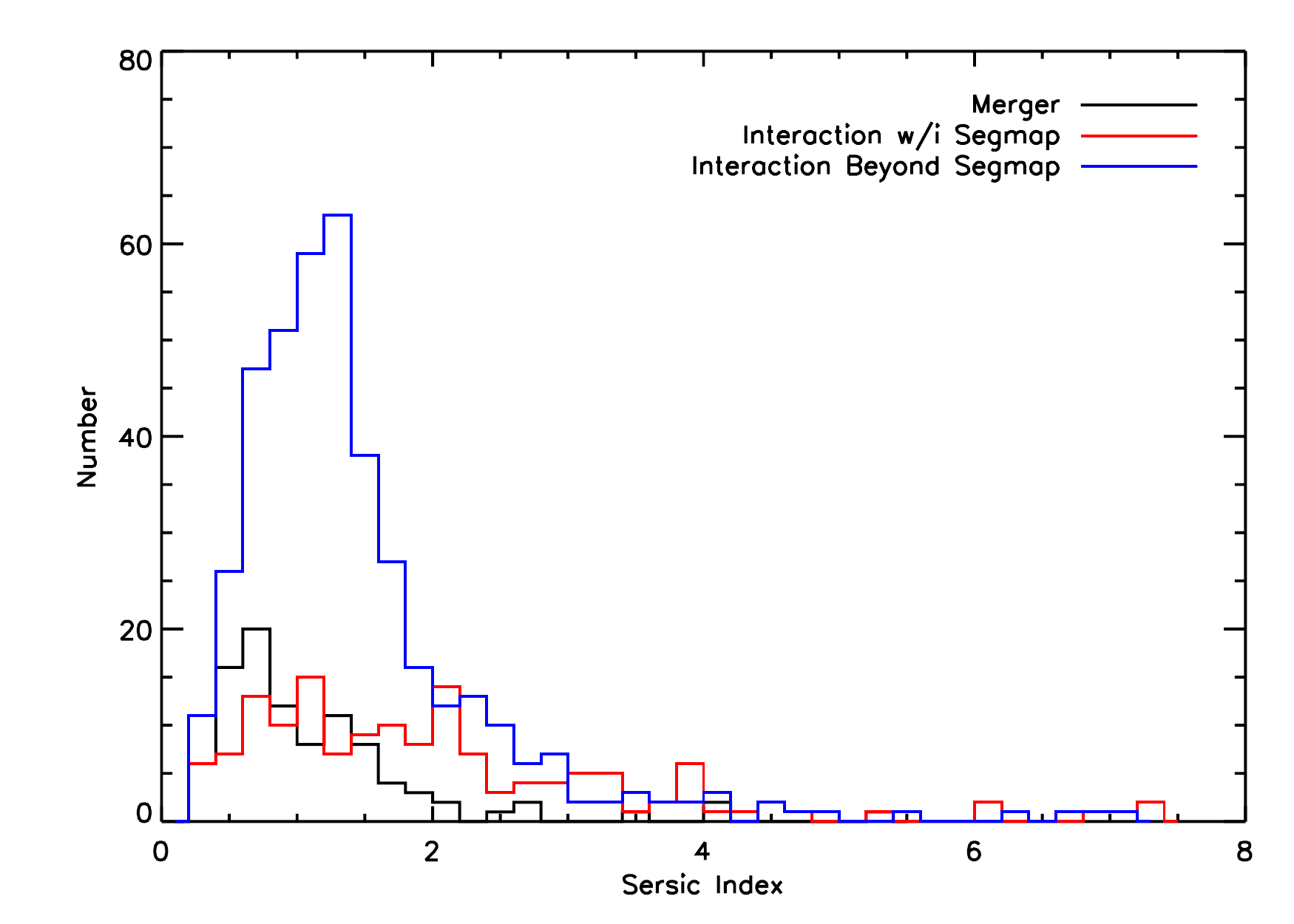}
\caption{Sersic index for visually classified galaxies in GOODS-S color coded by their main morphology class (top) and the interaction class (bottom). The main morphology classes are further refined by their relative diskiness/bulginess, into `Mostly Disk'  ($>2/3$ classifiers call it a disk and $<2/3$ classifiers call it a spheroid), `Mostly Spheroid' ($>2/3$ classifiers call it a spheroid and $<2/3$ classifiers call it a disk), `Disk + Spheroid' ($>2/3$ classifiers call it a disk and $>2/3$ classifiers call it a spheroid) and `Irregular' ($>2/3$ classifiers call it irregular, $<2/3$ classifiers call it a disk, and $<2/3$ classifiers call it a spheroid).}
\label{galfit}
\end{figure}

For the interaction classes in the bottom panel, we plot the distributions of Sersic indices for mergers, interactions within, and interactions beyond the segmentation map (in all cases, we use a threshold of $\geq 2/3$ classifiers). There is a difference between the distributions for these interaction classes, though slight. The mergers have the lowest mean Sersic index with $\langle n\rangle=1.15$. For the interactions beyond the segmentation map, the distribution is similar to the irregulars and disks in the top panel (with $\langle n\rangle=1.45$), which could be explained by the companion being distant enough that it does not affect the light distribution of the main galaxy by much. Finally, the interaction within the segmentation map shows the broadest distribution as well as the largest mean value ($\langle n\rangle=2.06$). Studies of the Sersic indices of merging galaxies in the local Universe have found a similar low mean Sersic index, similar to that of disks \citep{2013ApJ...768..102K}. This illustrates that the low Sersic index of mergers in our sample is not surprising that that the Sersic index itself is not a good measure of whether or not a galaxy is a merger.

To further explore how well our classifications separate disks and spheroids, we plot the positions of the sample on a colorÐcolor diagram (UVJ diagram: $U-V$ versus $V-J$) in Figure~\ref{uvj}. In this diagram, star forming galaxies follow a diagonal sequence with redder $V-J$ colors due to dust reddening while quiescent galaxies lie in a clump above the sequence with redder $U-V$ colors but bluer $V-J$ colors \citep[e.g.,][]{2007ApJ...655...51W,2009ApJ...700..799W,2009ApJ...691.1879W}. We split the sample into six different redshift bins and color-code the points by their main morphology class. Galaxies that are `Mostly Disks' separate cleanly from those that are  `Mostly Spheroids,' such that the disks fall onto the star forming sequence while the spheroids fall into the quiescent region (with some scatter). Galaxies classified as disk+spheroid mostly fall onto the star-forming sequence with some scatter up into the quiescent region while irregular galaxies mostly fall onto the star-forming sequence and are concentrated at blue $V-J$ colors. 

\begin{figure}
\epsscale{1.2}
\plotone{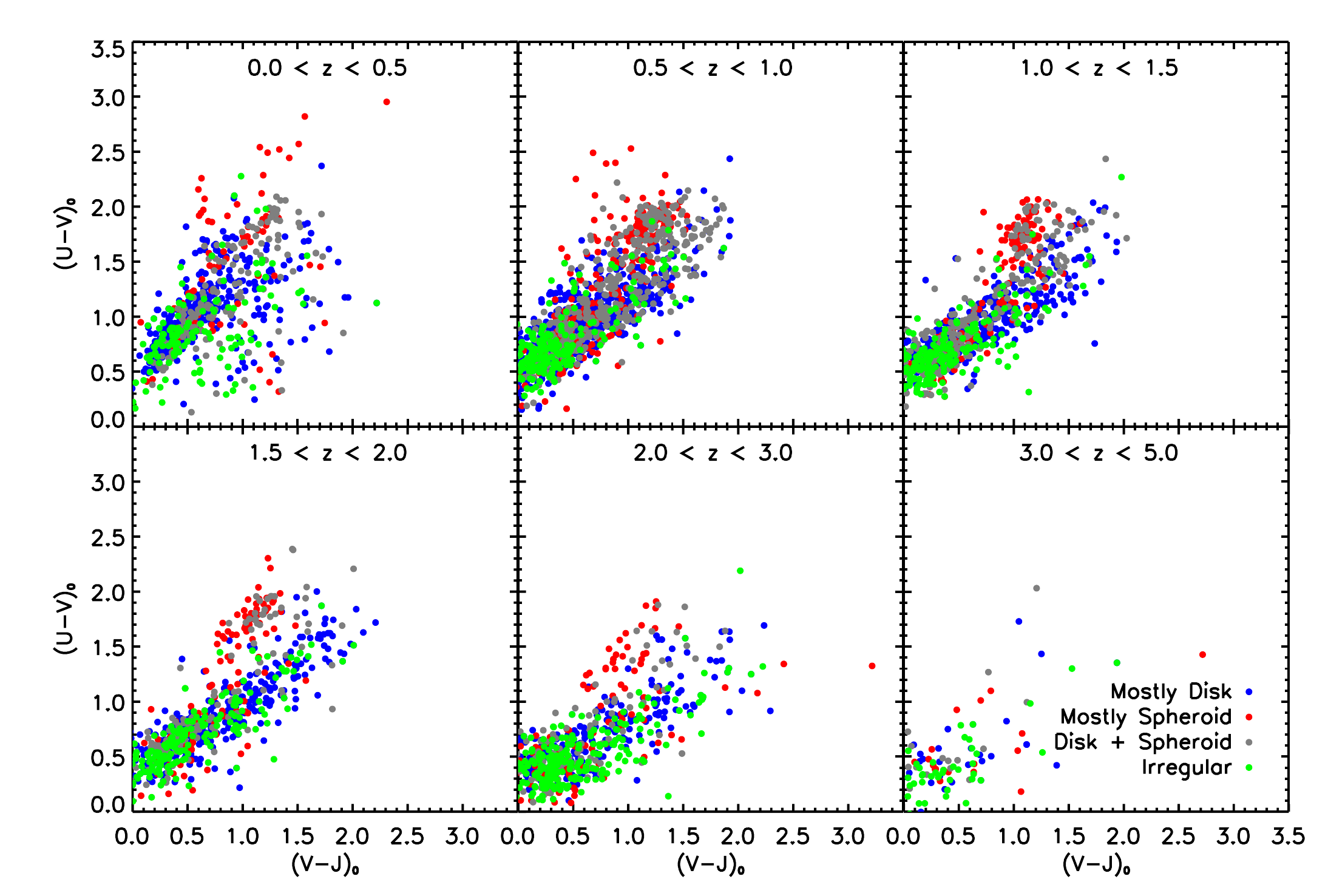}
\caption{UVJ diagram for our sample divided into six redshift bins and color coded by their dominant main morphology class (mostly disk, mostly spheroid, disk+spheroid, and irregular). Note that the disks and spheroids separate from each other as expected, the irregular galaxies do not occupy the quiescent region of the diagram, and the irregular galaxies are concentrated at blue $(V-J)_{0}$ colors. }
\label{uvj}
\end{figure}

\begin{figure*}
\epsscale{1}
\plotone{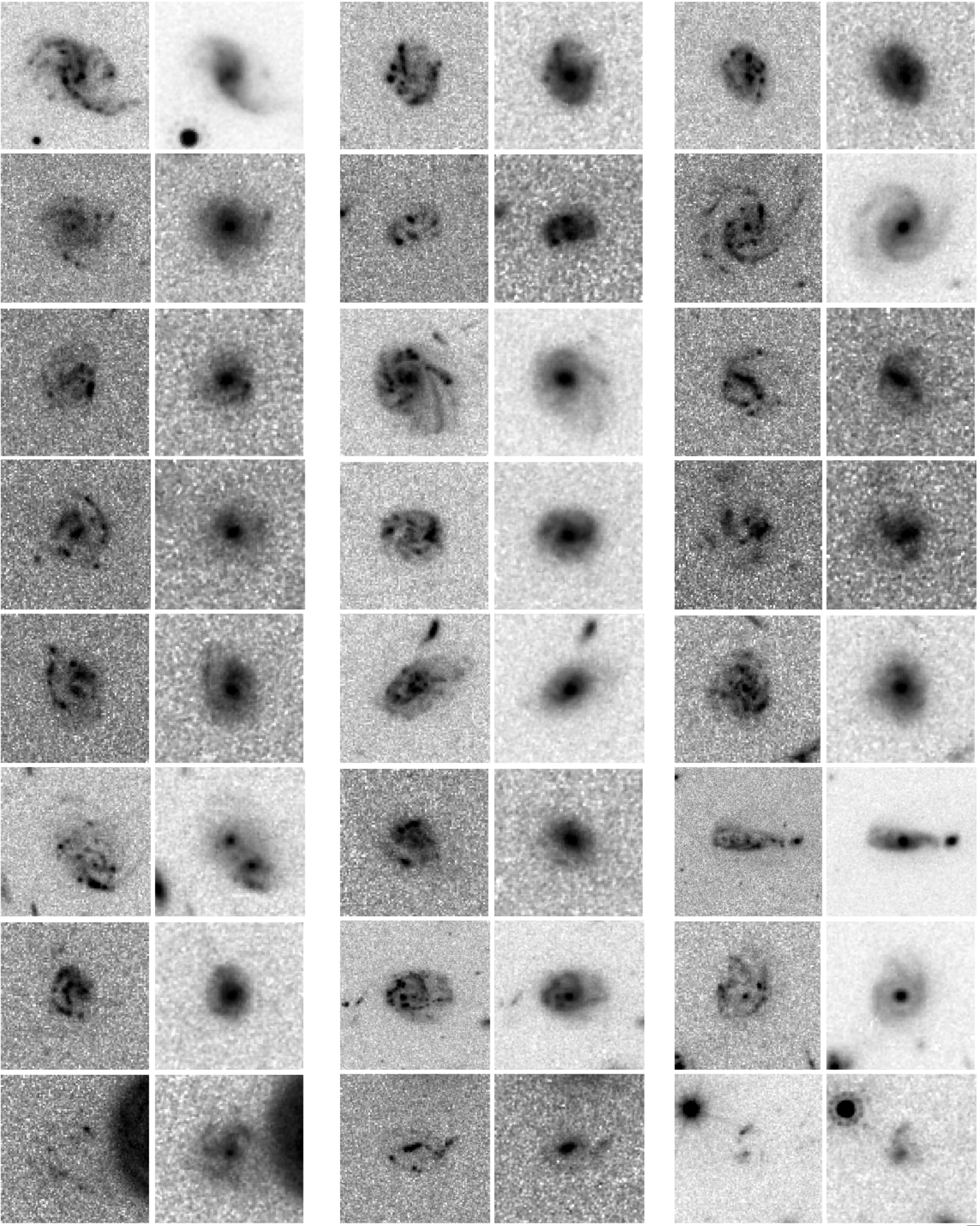}
\caption{Postage stamps of galaxies with a large morphological {\it k}-correction between the {\it V}-band (left, rest-frame UV) and {\it H}-band (right, rest-frame optical). The sizes of the stamps follow the prescription described in Section 2.3.}
\label{kcorrection}
\end{figure*}

\subsection{Morphological $K$-corrections}
    
Since the classifiers look at all of the bands at once, instead of individually, we cannot directly compare classifications between the different bands. However, we included a {\it k}-correction flag in the classification scheme so that the classifiers could mark cases that they think they would have classified differently in one or more of the other bands. This allows us to track and search for galaxies with large morphological {\it k}-corrections. Figure~\ref{kcorrection} contains a sample of objects that more than half of the classifiers flagged as being different in the {\it V}-band. This represents a total of 84 objects out of the full sample with $\langle z\rangle = 0.9$. Both the {\it V} and {\it H}-band stamps are shown side by side for comparison. Many of the objects that people mark as different are simply faint or undetected in the optical bands. However, most of the galaxies in Figure~\ref{kcorrection} have clumpy irregular light distributions in the optical and are more regular and smooth and often have more prominent bulges in the near-infrared.  For this reason, the classifiers are asked to look at the optical images when they select among the clumpiness/patchiness flags and in the near-infrared when choosing between bulge and disk dominated flags. These galaxies illustrate some of the pitfalls of basing classifications solely on the rest-frame UV light of high redshift galaxies -- some that seem highly irregular are actually normal looking disks in the rest-frame optical. Similarly, some galaxies that appear to be a single object in the rest-frame UV show up as pairs in the rest-frame optical.

\section{Summary}

We have presented an ambitious program to visually classify all galaxies in the five CANDELS fields down to $H<24.5$. Once completed, we expect to have detailed morphological classifications for over 50,000 galaxies spanning $0<z<4$. Our visual classification scheme was designed to cover a wide range of science goals and we have described each of its components in detail. With the publication of this paper, we release to the public all of the visual classifications in GOODS-S, including separate catalogs for the full field and deep region, with the deep region classified by five different people at each of three different depths. The remainder of the field (the `wide' portion and the ERS) have been classified by at least 3 people. Included in this data release is the calibration set of 200 galaxies that have been classified by all 65 classifiers. We also release the software for the Perl/Tk GUI that we developed to classify galaxies.

We have made a number of comparisons among the classifiers to test for internal consistency and find that the level of agreement is dependent on both the brightness of the galaxy ({\it H}-band magnitude) and the type of galaxy itself. The classifications are also consistent across the different depths for all but the faintest galaxies. Our detailed calibration set of galaxies illustrates the complexity of galaxy morphology for these faint objects and indicates the necessity of having multiple classifiers for each object. A comparison of our visual classifications with Sersic indices measured by GALFIT shows broad agreement, where galaxies classified as disks tend to have lower Sersic indices and galaxies classified as spheroids tend to have higher Sersic indices. We also find that the colors of our morphologically selected galaxies are consistent with what we would expect for their galaxy types, such that disks are on the star forming sequence and spheroids are mostly quiescent. Irregular galaxies and galaxy mergers and interactions are the hardest to identify in our sample and tend to show the highest level of disagreement. A future paper (J. S. Kartaltepe et al. in prep) will investigate galaxy mergers and interactions in these classifications in more detail.

\newpage

\acknowledgments
Support for this work was provided by NASA through Hubble Fellowship grant \# HST-HF-51292.01A awarded by the Space Telescope Science Institute, which is operated by the Association of Universities for Research in Astronomy, Inc., for NASA, under contract NAS 5-26555. Support for Program number HST-GO-12060  was provided by NASA through a grant from the Space Telescope Science Institute, which is operated by the Association of Universities for Research in Astronomy, Incorporated, under NASA contract NAS5-26555. O'Leary and Blancato were supported by the NOAO/KPNO Research Experiences for Undergraduates (REU) Program with two grants from the National Science Foundation Research Experiences for Undergraduates Program (AST-0754223, AST-1262829).

{\it Facilities:} \facility{HST (WFC3)}



\begin{deluxetable}{lcccccccccc}
  \tablewidth{0pt}
  \tablecaption{CANDELS: GOODS-S Visual Classification Catalog Data Release \label{list}}
\tablehead{\colhead{Depth} & \colhead{Type\tablenotemark{a}} &  \colhead{Area} &  \colhead{\# of entries} & 
              \colhead{\# of objects} 
              }
\startdata
2-epoch & raw          & GOODS-S deep+wide+ERS & 42695 & 7634 \\
2-epoch & fractional & GOODS-S deep+wide+ERS & 7634 & 7634 \\
2-epoch & raw          & GOODS-S deep                    & 25059 & 2534 \\
2-epoch & fractional & GOODS-S deep                    & 2534 & 2534 \\
4-epoch & raw          & GOODS-S deep                    & 12670 & 2534 \\
4-epoch & fractional & GOODS-S deep                    & 2534 & 2534 \\
10-epoch & raw          & GOODS-S deep                    & 13477 & 2534 \\
10-epoch & fractional & GOODS-S deep                    & 2534 & 2534 
\enddata
\tablenotetext{a}{The raw classification catalogs follow the format of Table~\ref{raw} while the fractional catalogs follow the format of Table~\ref{fraction}.}
\end{deluxetable}

\begin{deluxetable}{lcccccccccc}
  \tablewidth{0pt}
  \tabletypesize{\scriptsize}
  \tablecaption{CANDELS: GOODS-S 2-Epoch Depth Raw Visual Classification Catalog \label{raw}}
  \setlength{\tabcolsep}{0.05in}
\tablehead{\colhead{ID} & \colhead{RA} & \colhead{Dec} & 
              \colhead{Spheroid\tablenotemark{a}} & \colhead{Disk} & \colhead{Irregular} & 
              \colhead{Compact} & \colhead{Unclassifiable} & \colhead{Interaction Class} & \colhead{Classifier} & \colhead{comments}
              }
\startdata
 GDS\_deep2\_10000  &  53.054728 & -27.769708 & 0 & 0 & 0 & 1 & 0 & 0 & 1 & \nodata \\
 GDS\_deep2\_10000  &  53.054728 & -27.769708 & 0 & 0 & 0 & 1 & 0 & 0 & 2 & \nodata     \\
 GDS\_deep2\_10000  & 53.054728 & -27.769708 & 1 & 0 & 0 & 0 & 0 & 0 & 3 & \nodata   \\
 GDS\_deep2\_10000  & 53.054728 & -27.769708 & 0 & 0 & 0 & 1 & 0 & 0 & 4 & \nodata   \\
 GDS\_deep2\_10000  & 53.054728 & -27.769708 & 1 & 0 & 0 & 0 & 0 & 0 & 5 & \nodata   
\enddata
\tablenotetext{}{Notes.--- Table~\ref{raw} is published in its entirety in the online edition of the article. A portion is shown here for guidance regarding its form and content.}
\tablenotetext{a}{All of the classifications discussed in this paper are presented in the online version of the paper. Here, only the main morphology class and interaction class columns are shown as an example.}
\end{deluxetable}

\begin{deluxetable}{lcccccccccc}
  \tablewidth{0pt}
  \tabletypesize{\scriptsize}
  \tablecaption{CANDELS: GOODS-S 2-Epoch Depth Fractional Visual Classification Catalog \label{fraction}}
  \setlength{\tabcolsep}{0.05in}
\tablehead{\colhead{ID} & \colhead{RA} & \colhead{Dec} &  \colhead{\#\tablenotemark{a}} &
              \colhead{Spheroid\tablenotemark{b}} & \colhead{Disk} & \colhead{Irregular} & 
              \colhead{Compact} & \colhead{Unclassifiable} & \colhead{Merger} 
              }
\startdata
 GDS\_deep2\_4407  & 53.0746 &    -27.8473 & 5 & 0.6 & 0.6 & 0.6 & 0.0 & 0.0 & 0.0  \\
 GDS\_deep2\_4418  & 53.1035  &   -27.8473 & 5 & 0.6 & 1.0 & 0.2 & 0.0 & 0.0 & 0.0   \\
 GDS\_deep2\_4420  & 53.0902   &  -27.8479 & 5 & 0.2 & 1.0 & 0.6 & 0.0 & 0.0 & 0.0    \\
 GDS\_deep2\_4422 &  53.0758   &  -27.8466 & 5 & 1.0 & 0.4 & 0.0 & 0.2 & 0.0 & 0.0     \\
 GDS\_deep2\_4423 & 53.0915    & -27.8468  & 5 & 0.4 & 0.2 & 0.6 & 0.0 & 0.0 & 0.0     
\enddata

\tablenotetext{}{Notes.--- Table~\ref{fraction} is published in its entirety in the online edition of the article. A portion is shown here for guidance regarding its form and content.}
\tablenotetext{a}{Number of people that classified this object.}
\tablenotetext{b}{All of the classifications discussed in this paper are presented in the online version of the paper. Here, only the main morphology class and interaction class columns are shown as an example.}
\end{deluxetable}


\end{document}